\documentclass{aa}  
\usepackage{color}
\usepackage{graphicx}
\usepackage{lscape}
\usepackage{natbib}
\usepackage[varg]{txfonts}
\usepackage{url}
\usepackage{xspace}
\bibpunct{(}{)}{;}{a}{}{,}
\def\etal{{et\,al.}\ }
\def\md{log($\dot{M}/M_{\sun} {\rm yr}^{-1}$)\,}
\newcommand{\Teff}{$T\mathrm{\hspace*{-0.4ex}_{eff}}$\,}
\newcommand{\logg}{$\log\,g$\hspace*{0.5ex}}

\def\hszwei{HS\,2115+1148}
\def\henull{HE\,0504--2408}
\def\hsnull{HS\,0713+3958}
\def\hsx{HS\,0727+6003}
\def\bd{BD$+10^\circ 2179$}
\def\re{RE\,0503--289}
\begin{document}

\title{Metal abundances in hot white dwarfs with signatures of a superionized wind}

\author{
K\@. Werner\inst{1} \and 
        T\@. Rauch\inst{1} \and
        J\@. W\@. Kruk\inst{2} 
}

\institute{Institute for Astronomy and Astrophysics, Kepler Center for Astro and
Particle Physics,  Eberhard Karls University, Sand~1, 72076
T\"ubingen, Germany\\ \email{werner@astro.uni-tuebingen.de}\and
           NASA Goddard Space Flight Center, Greenbelt, MD\,20771, USA}

\date{Received 8 August 2017 / Accepted 16 October 2017}

\authorrunning{K. Werner \etal}
\titlerunning{Metal abundances in hot white dwarfs with signatures of a superionized wind}

\abstract{About a dozen  hot white dwarfs with effective temperatures
  \Teff = 65\,000\,K -- 120\,000\,K exhibit unusual absorption
  features in their optical spectra. These objects were tentatively
  identified as Rydberg lines of ultra-high excited metals in
  ionization stages {\sc v--x}, indicating line formation in a dense
  environment with temperatures near $10^6$\,K. Since some features
  show blueward extensions, it was argued that they stem from a
  superionized wind. A unique assignment of the lines to particular
  elements is not possible, although they probably stem from C, N, O,
  and Ne.  To further investigate this phenomenon, we analyzed the
  ultraviolet spectra available from only three stars of this group;
  that is, two helium-rich white dwarfs, \henull\ and \hsnull\ with
  spectral type DO, and a hydrogen-rich white dwarf, \hszwei\ with
  spectral type DAO. We identified light metals (C, N, O, Si, P, and
  S) with generally subsolar abundances and heavy elements from the
  iron group (Cr, Mn, Fe, Co, Ni) with solar or oversolar
  abundance. The abundance patterns are not unusual for hot WDs and
  can be interpreted as the result of gravitational settling and
  radiative levitation of elements. As to the origin of the ultra-high
  ionized metals lines, we discuss the possible presence of a
  multicomponent radiatively driven wind that is frictionally heated.}


\maketitle
%

\begin{table*}[t]
\begin{center}
\caption{Observation log of our three program stars (first three
  objects listed) and an additional white dwarf discussed in this
  paper.\tablefootmark{a} }
\label{tab:obs} 
\small
\begin{tabular}{cccccccccc}
\hline 
\hline 
\noalign{\smallskip}
Star    & Type & Instrument & Dataset   & Grating  & $R$        &$\lambda$/\AA& $t_{\rm exp}$/s&Date&PI  \\
\hline 
\noalign{\smallskip}
\hszwei & DAO  & HST/GHRS   & Z3GT0104T & G140L    & 1700--2200 & 1150--1435  & 4243 & 1996-10-23 &Werner\\
        &      & HST/GHRS   & Z3GT0105T & G140L    & 1700--2200 & 1480--1770  & 5440 & 1996-10-23 &Werner\\
        &      &  FUSE       & C0960101000&         & 20000      & 915--1188   & 9370 & 2002-07-02 &Finley\\
        &      &  FUSE       & C0960102000&         & 20000      & 915--1188   & 6441 & 2004-05-59 &Finley\\
\noalign{\smallskip}
\henull & DO   & HST/GHRS   & Z2WX0104T & G140L    & 1700--2200 & 1150--1435  & 1414 & 1995-09-28 &Werner\\
        &      &  HST/GHRS   & Z2WX0105T & G140L    & 2300--2700 & 1480--1770  & 2611 & 1995-09-28 &Werner\\
        &      &  HST/GHRS   & Z2WX0106T & G270M    & 30000      & 2957--3000  & 2394 & 1995-09-28 &Werner\\
        &      &  FUSE       & A0010101000&         & 20000      & 915--1188   & 6620 & 2001-12-03 &Werner\\
\noalign{\smallskip}
\hsnull & DO   &  HST/GHRS   & Z2WX0204T & G140L    & 1700--2200 & 1150--1435  & 4243 & 1995-09-22 &Werner\\
        &      &  HST/STIS\tablefootmark{b}&O63Z01010& G750L    & 530--1040  & 5240--10270 & 2405 & 2001-01-19 &Werner\\
        &      &  FUSE       & A0010201000&         & 20000      & 915--1188   & 5249 & 2000-03-15& Werner\\
        &      &  FUSE       & A0010202000&         & 20000      & 915--1188   & 9812 & 2000-11-11 &Werner\\
        &      &  FUSE       & S6011901000&         & 20000      & 915--1188   & 8933 & 2002-02-12& Friedman\\
\noalign{\smallskip}
\hsx    & DO   &  FUSE       & Z9031001000&         & 20000      & 915--1188   &20556 & 2003-02-01& Dupuis\\

\noalign{\smallskip} \hline
\end{tabular} 
\tablefoot{\tablefoottext{a}{All FUSE datasets observed with LWRS
    aperture. Resolving power is $R$. Exposure time is t$_{\rm
      exp}$.}\tablefoottext{b}{Observation of the cool companion.}} 
\end{center}
\end{table*}

\begin{figure*}[t]
 \centering  \includegraphics[width=1.0\textwidth]{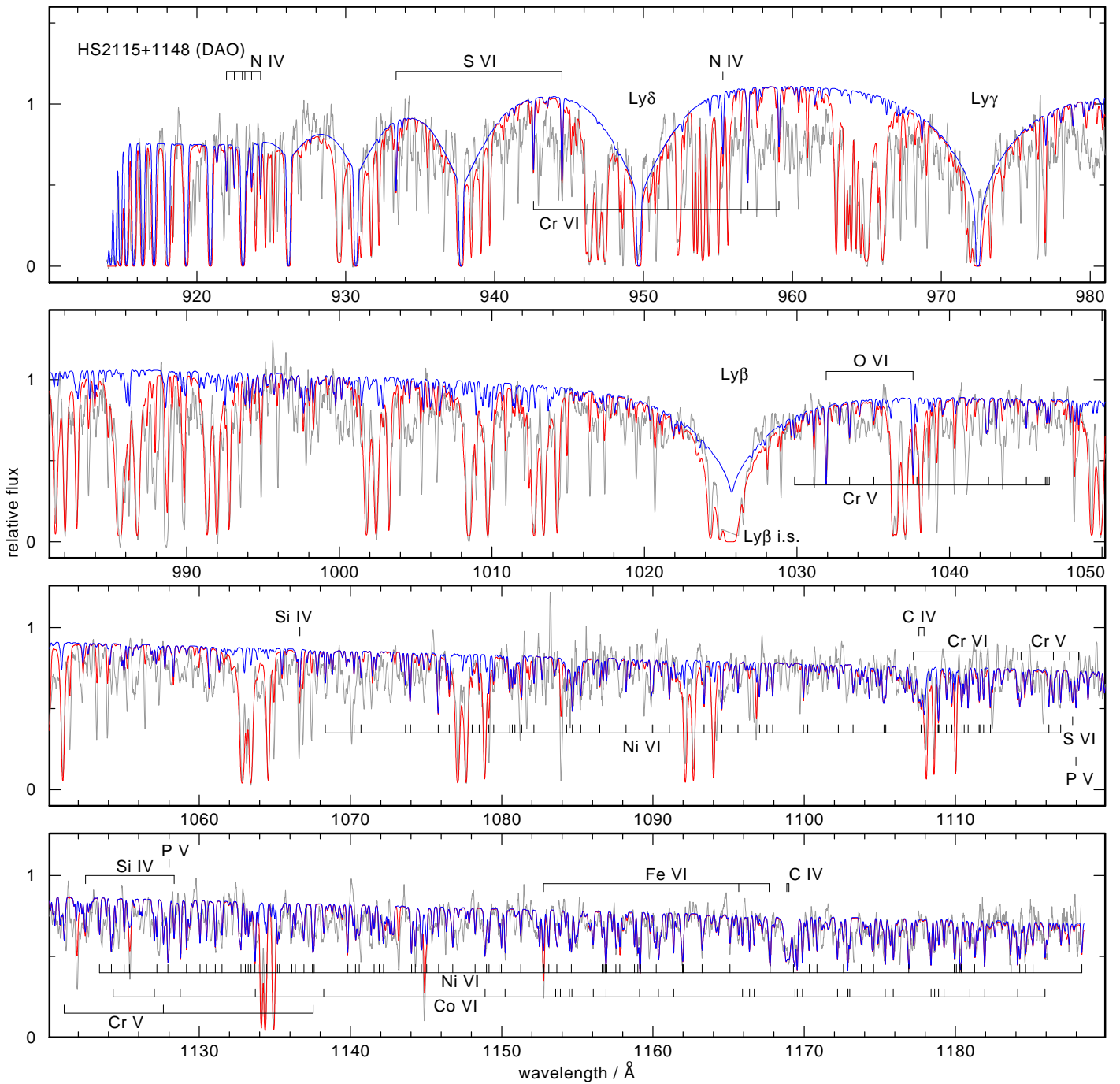}
  \caption{FUSE spectrum of the DAO \hszwei\ (gray line) compared to a
    photospheric model spectrum (blue line; \Teff = 80\,000\,K, \logg = 7) with the finally
    adopted parameters as listed in
    Table\,\ref{tab:stars}. The same model
    including interstellar absorption lines is overplotted in red. Prominent spectral lines are
    identified.}
\label{fig:hs2115_fuse}
\end{figure*}

\begin{figure*}[t]
 \centering  \includegraphics[width=1.0\textwidth]{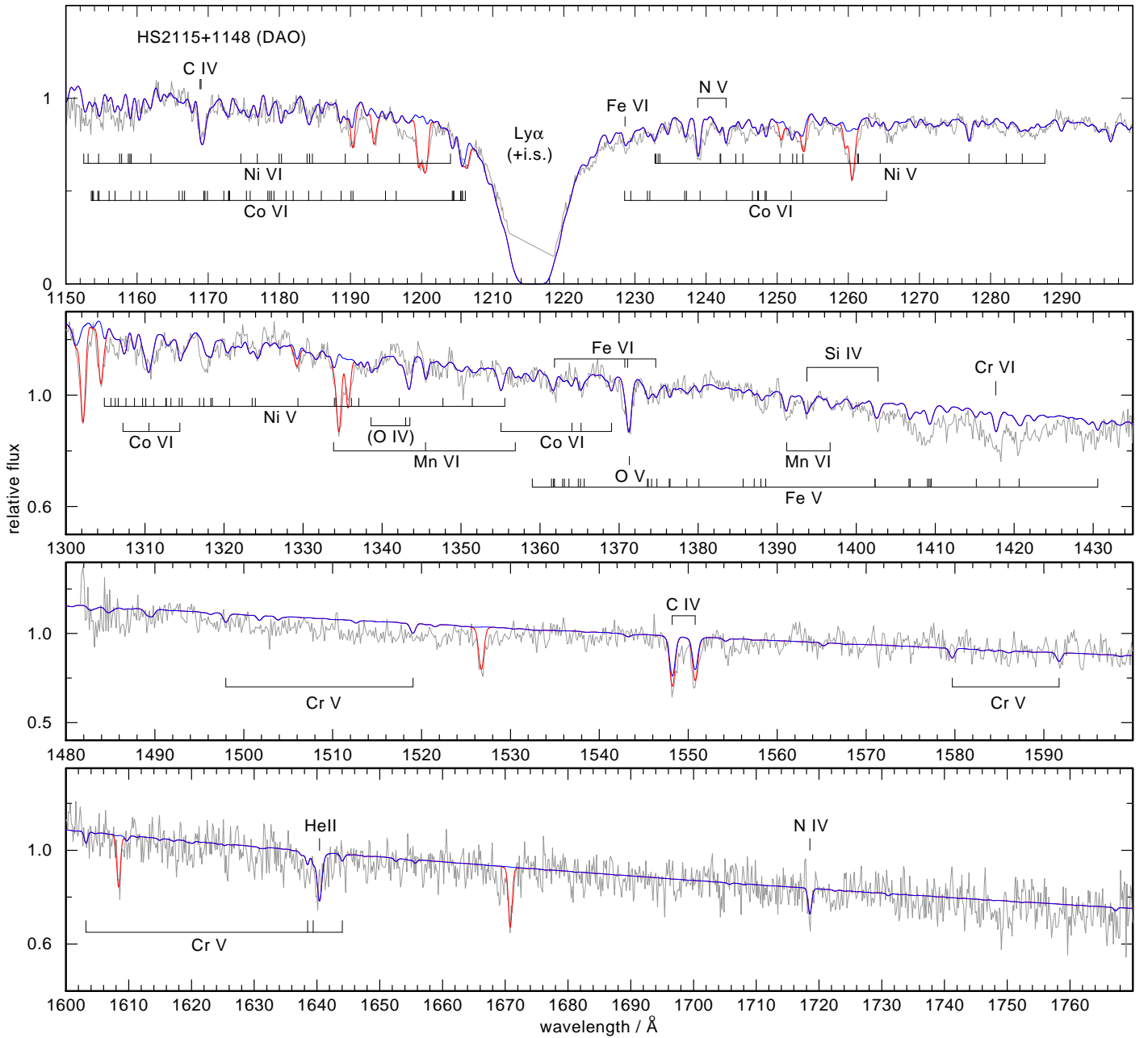}
  \caption{HST\ spectrum of the DAO \hszwei, similar to Fig.\,\ref{fig:hs2115_fuse}. Line identifications enclosed in brackets
    denote non-detections in the observations. Model: \Teff = 80\,000\,K, \logg = 7.}
\label{fig:hs2115_hst}
\end{figure*}
\
\begin{figure*}[t]
 \centering  \includegraphics[width=1.0\textwidth]{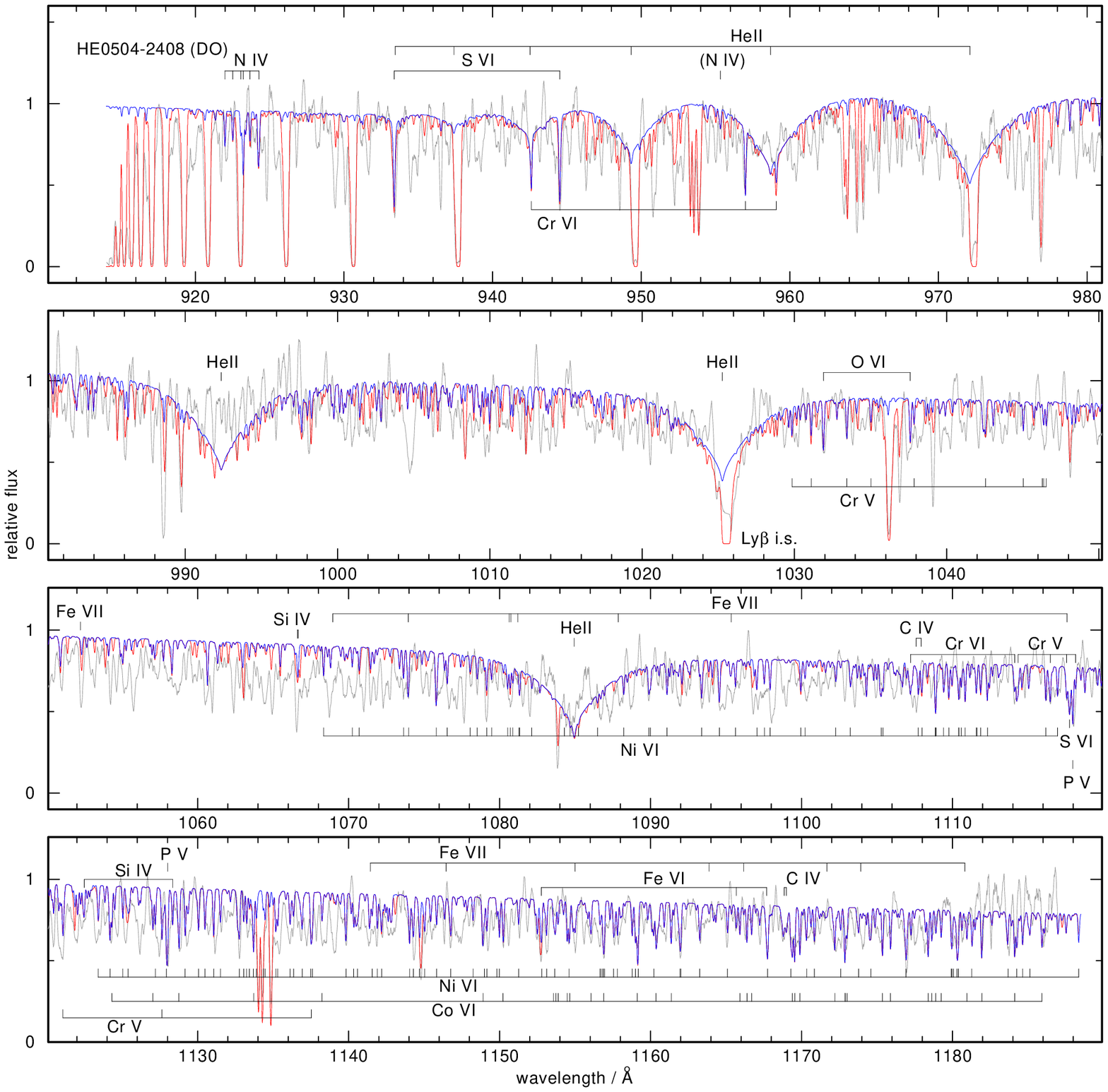}
  \caption{FUSE\ spectrum of the DO \henull, similar to Fig.\,\ref{fig:hs2115_fuse}. Model: \Teff = 85\,000\,K, \logg = 7.}
\label{fig:he0504_fuse}
\end{figure*}

\begin{figure*}[t]
 \centering  \includegraphics[width=1.0\textwidth]{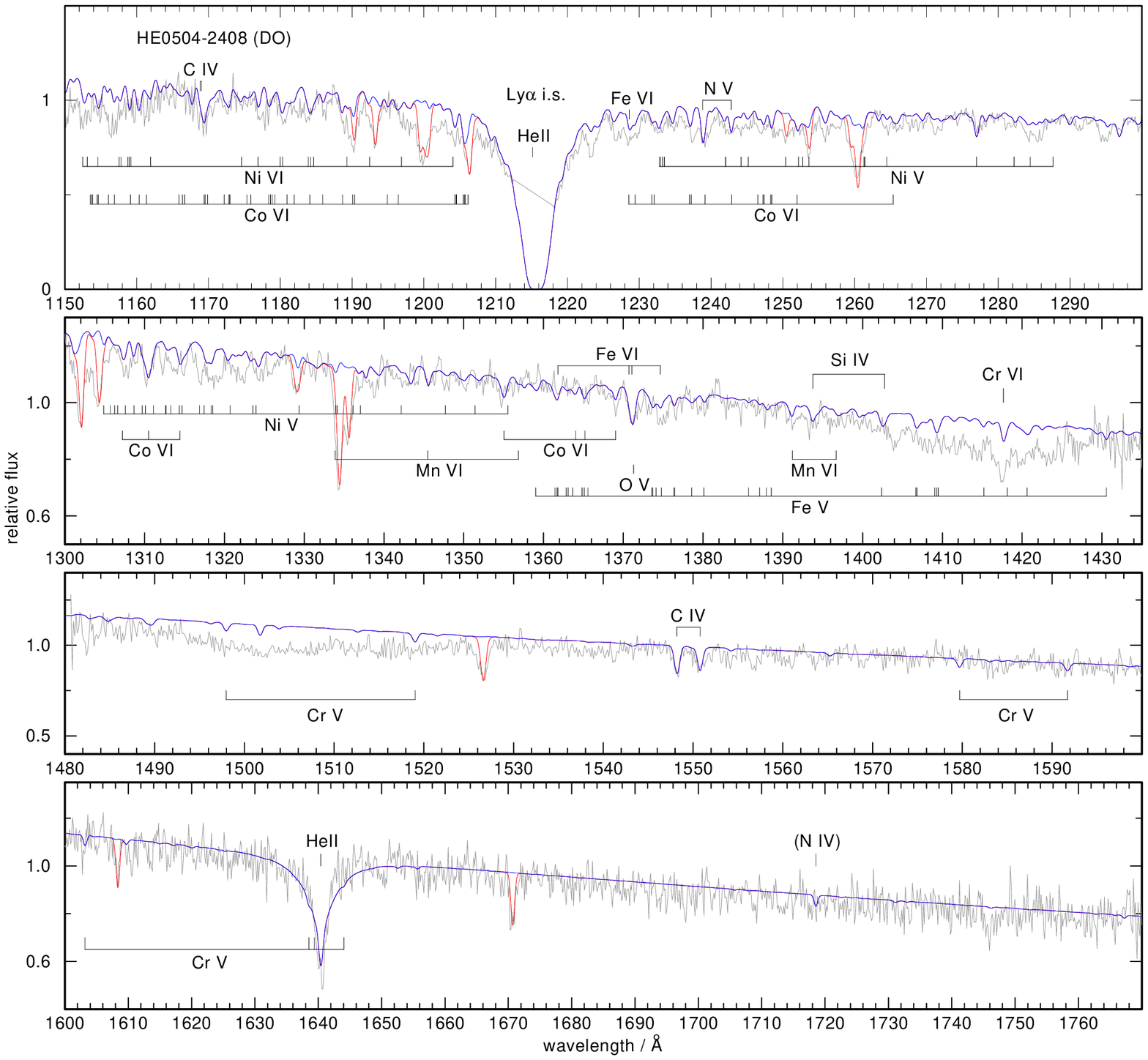}
  \caption{HST\ spectrum of the DO \henull, similar to Fig.\,\ref{fig:hs2115_fuse}. Model: \Teff = 85\,000\,K, \logg = 7.}
\label{fig:he0504_hst}
\end{figure*}

\begin{figure*}[t]
 \centering  \includegraphics[width=1.0\textwidth]{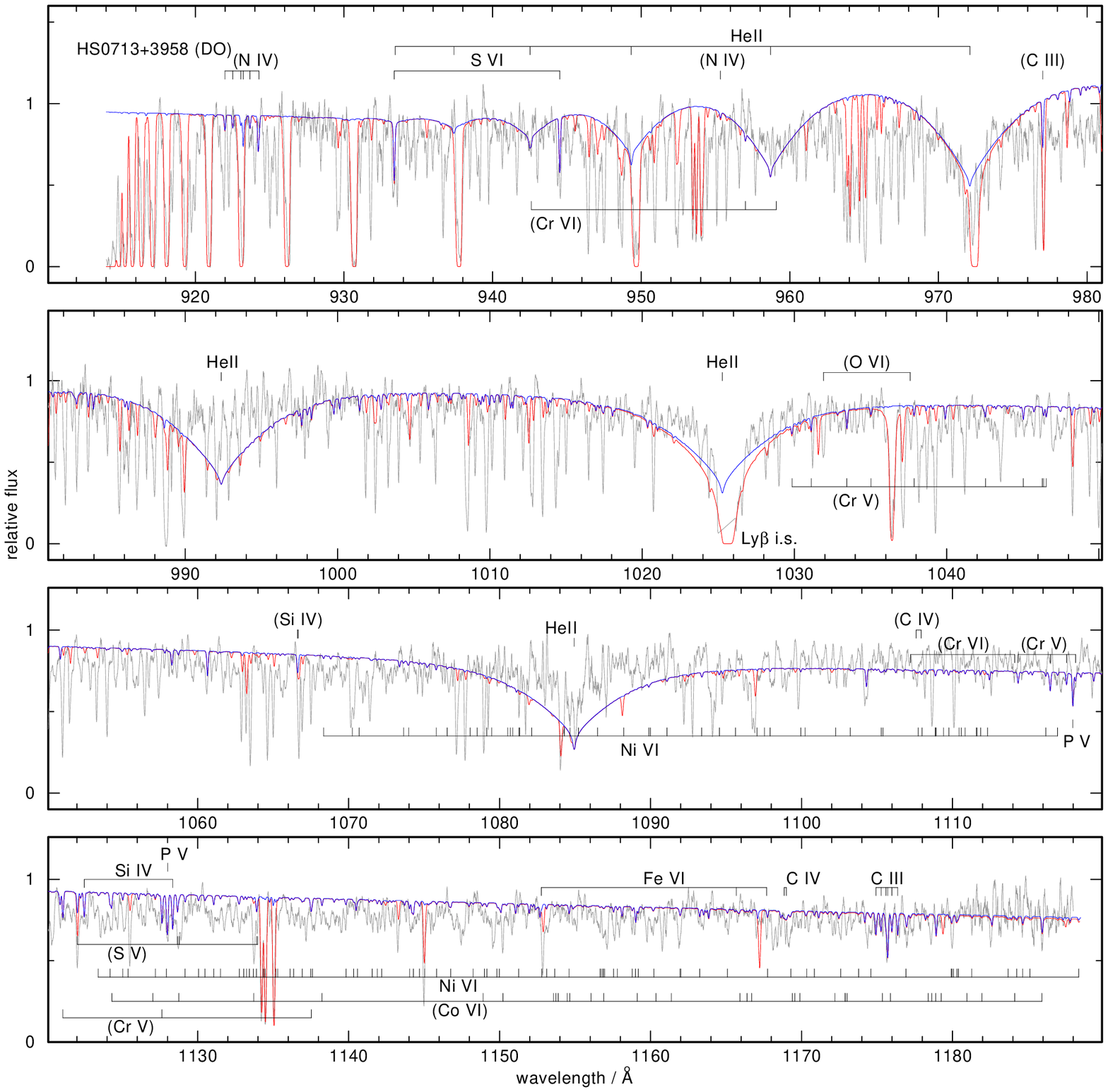}
  \caption{FUSE spectrum of the DO \hsnull, similar to Fig.\,\ref{fig:hs2115_fuse}. Model: \Teff = 65\,000\,K, \logg = 7.5.}
\label{fig:hs0713_fuse}
\end{figure*}

\begin{figure*}[t]
 \centering  \includegraphics[width=1.0\textwidth]{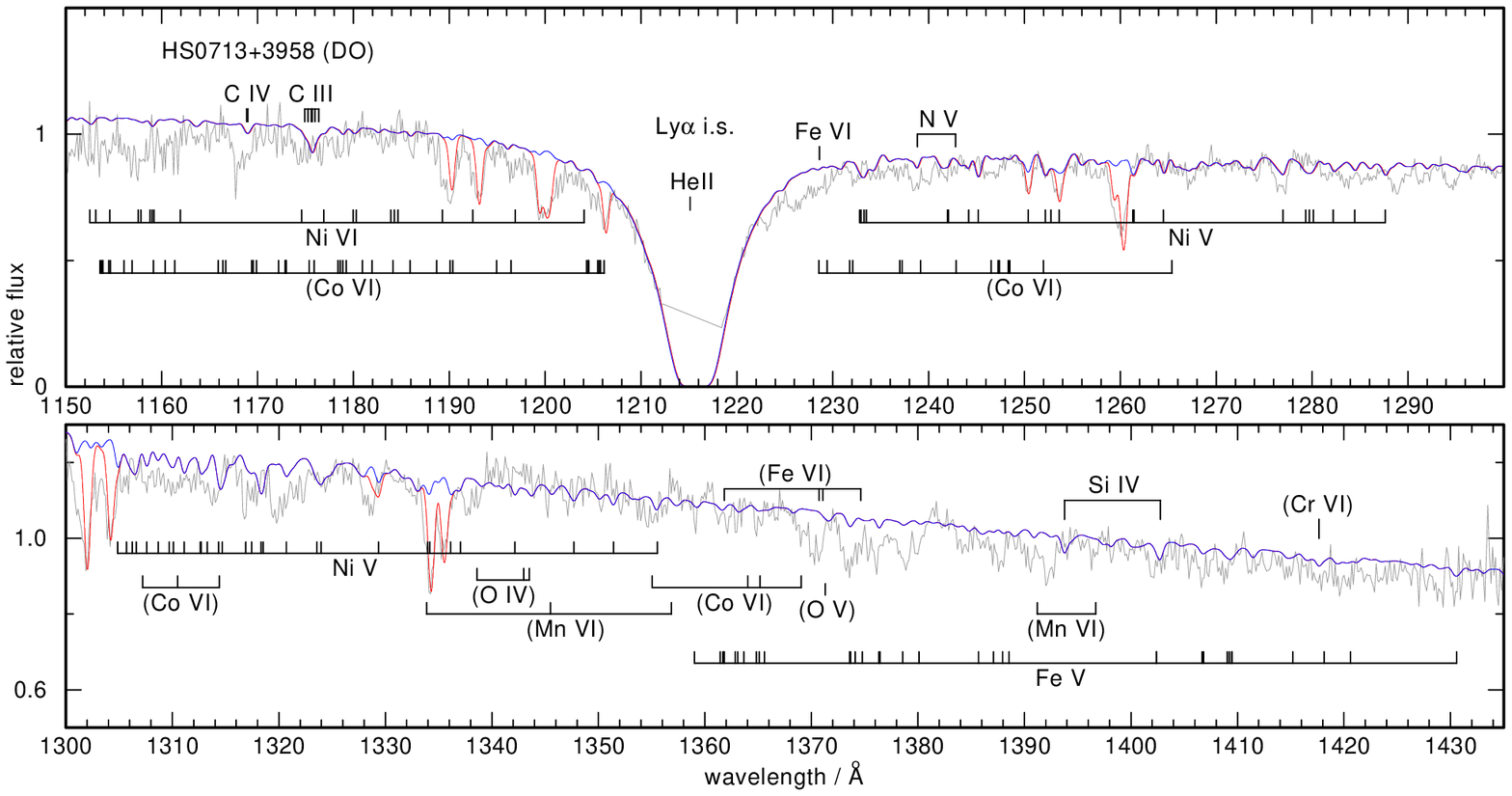}
  \caption{HST spectrum of the DO \hsnull, similar to
    Fig.\,\ref{fig:hs2115_fuse}. Model: \Teff =
    65\,000\,K, \logg = 7.5.}
\label{fig:hs0713_hst}
\end{figure*}

\section{Introduction}
\label{intro}

More than two decades ago, the discovery of two hot helium-rich (DO)
white dwarfs (\henull, \hsnull; effective temperature \Teff $\approx$
70\,000\,K) with ultra-high excitation (\emph{uhe}) absorption lines
in optical spectra was announced \citep{1995A&A...293L..75W}. The
newly detected lines were assigned to extremely high ionization stages
of the CNO elements and neon (\ion{C}{v}, \ion{C}{vi}/\ion{N}{vi},
\ion{N}{vii}/\ion{O}{vii}, \ion{O}{viii}, \ion{Ne}{ix}, \ion{and
  Ne}{x}). The ionization energies, however, require temperatures on
the order $10^6$\,K. Some lines exhibit an asymmetric profile shape
suggesting their formation in a stellar wind, hence the alternative
designation of these stars as ``hot-wind DOs''. Since then, no
progress has been made to explain the origin of these line
features. However, this phenomenon turned out to be not uncommon among
DO white dwarfs. In total, 9 out of about 70 known hot DOs are
affected
\citep{1995A&A...303L..53D,2006A&A...454..617H,2014A&A...572A.117R,2014A&A...564A..53W}. In
addition, one DAO white dwarf \citep[\hszwei,][]{1995A&A...303L..53D}
and one PG1159 star \citep{2006A&A...454..617H} were found to exhibit
\emph{uhe} lines.

The determination of the basic photospheric parameters (\Teff\ and
surface gravity $g$) of these stars is hampered by an obviously
related phenomenon in the optical spectra, namely abnormally deep
\ion{He}{ii} lines, in the case of the DOs and PG1159 star, and Balmer
lines, in the case of the DAO. In this paper we present the first
analysis of ultraviolet (UV) spectra that are available for only three
\emph{uhe} stars; i.e., the two DOs and the DAO mentioned above. We
performed the observations with the Goddard High Resolution
Spectrograph (GHRS) aboard the \emph{Hubble Space Telescope} (HST) and
the \emph{Far Ultraviolet Spectroscopic Explorer} (FUSE) and use
additional archival data. The primary aim of this analysis is to
constrain the stellar effective temperatures by several ionization
balances and determine metal abundances. Further clues concerning the
\emph{uhe} phenomenon are expected as well.

We introduce the program stars in Sect.\,\ref{sect:programstars} and
present the UV observations in Sect.\,\ref{sect:observations}. Model
atmospheres and atomic data for the spectral analysis are described in
Sect.\,\ref{sect:models}. Spectral line fitting procedures and results
are reported in Sect.\,\ref{sect:results}. The paper is concluded with
a summary and discussion of the results
(Sect.\,\ref{sect:discussion}.)

\section{Program stars}
\label{sect:programstars}

Our program stars are those three WDs exhibiting \emph{uhe} features,
for which FUSE spectra and HST spectra with sufficient resolution
spectra are available. For another DO, HS\,2027+0651
\citep{1995A&A...303L..53D}, only low resolution (1.2\,\AA\ and less)
HST/STIS data exist, and hence, it was not included in our
study. Still another \emph{uhe} DO discovered by
\citet{1995A&A...303L..53D}, \hsx, has not been observed with HST, and
hence, we do not analyze this object in detail either. However,
\hsx\  has an archival, unpublished FUSE spectrum, which we will
shortly discuss.

\subsection{DO white dwarfs \henull\ and \hsnull}\label{sect:dos}

A rough estimate of \Teff = 70\,000\,K and \logg = 7.5 followed from
optical spectra of both stars \citep{1995A&A...293L..75W}.
\hsnull\ was reassessed by \citet{2014A&A...566A.116R} using newly
available optical SDSS spectra and a grid of He+C NLTE model
atmospheres, and \Teff = 80\,000$\pm$10\,000\,K and \logg =
7.75$\pm$0.5 was derived. The detection of \ion{He}{i}
$\lambda$5876\,\AA\ served as a \Teff\ constraint and also supports
our present analysis. Infrared photometry revealed that \hsnull\ has a
cool companion with a spectral type of late K that is separated about
1'' from the primary \citep{1997fbs..conf..207N}.

\subsection{\hszwei\ (DAO)}\label{sect:dao}

In their discovery paper, \citet{1995A&A...303L..53D} derived \Teff =
67\,000\,K and \logg = 6.9 from the Balmer lines, putting emphasis on
the highest observed series member (H$\delta$) and noting a severe
Balmer-line problem: H$\alpha$--H$\gamma$ are much deeper in the
observation than in the model with the given parameters (H+He NLTE
models). Similar problems with other DAO stars
\citep{1994A&A...285..603N} were solved by the inclusion of C, N, and
O opacities \citep{1996ApJ...457L..39W}, however, in this particular
case the problem remained. \citet{2010ApJ...720..581G} derived \Teff =
62\,230\,K, \logg = 7.76 from Balmer-line fitting using NLTE models
including CNO as the only metals and assuming solar abundances. The
inclusion of CNO was used as a proxy for the presence of metals to
mitigate the Balmer-line problem. The two analyses arrived at similar
helium abundances of 0.63\,\% and 0.43\,\% mass fraction, which were
derived from a weak \ion{He}{ii} $\lambda$4686\,\AA\ line.

Compared to the two DOs, the \emph{uhe} lines in this DAO are much
less prominent.

\begin{table}[t]
\begin{center}
\caption{Finally adopted parameters for the three program stars.\tablefootmark{a} }
\label{tab:stars} 
\tiny
\begin{tabular}{rrrrrr}
\hline 
\hline 
\noalign{\smallskip}
           & HS                  & HE                   & HS\\
           & 2115+1148           & 0504$-$2408          & 0713+3958          &  Sun\tablefootmark{b}\\ 
\hline 
\noalign{\smallskip}
Type      &  DAO                 & DO                   & DO \\
\Teff/\,K & 80\,000              & 85\,000              & 65\,000 \\
\logg     & 7.0                  & 7.0                  & 7.5 \\
\noalign{\smallskip}
H         & 0.99                 & $-$                  & $-$                  & 0.74\\
He        & $ 1.0 \times 10^{-3}$ & 0.995                & 0.999                & 0.25\\
C         & $ 3.0 \times 10^{-4}$ & $ 3.0 \times 10^{-5}$ & $ 7.0 \times 10^{-6}$ & $2.4 \times 10^{-3}$\\
N         & $ 3.0 \times 10^{-6}$ & $<1.0 \times 10^{-6}$ & $<3.0 \times 10^{-7}$ & $6.9 \times 10^{-4}$\\
O         & $ 7.0 \times 10^{-5}$ & $ 1.0 \times 10^{-5}$ & $<3.0 \times 10^{-6}$ & $5.7 \times 10^{-3}$\\
Si        & $ 3.0 \times 10^{-5}$ & $ 3.0 \times 10^{-5}$ & $ 5.0 \times 10^{-6}$ & $6.6 \times 10^{-4}$\\
P         & $ 1.0 \times 10^{-6}$ & $ 1.0 \times 10^{-5}$ & $ 1.5 \times 10^{-7}$ & $5.8 \times 10^{-6}$\\
S         & $ 5.0 \times 10^{-6}$ & $ 5.0 \times 10^{-5}$ & $ 1.0 \times 10^{-5}$ & $3.1 \times 10^{-4}$\\
Cr        & $ 1.3 \times 10^{-3}$ & $ 1.3 \times 10^{-3}$ & $<1.3 \times 10^{-4}$ & $1.7 \times 10^{-5}$\\
Mn        & $ 3.0 \times 10^{-5}$ & $ 3.0 \times 10^{-5}$ & $-$                  & $1.1 \times 10^{-5}$\\
Fe        & $ 1.3 \times 10^{-3}$ & $ 1.3 \times 10^{-3}$ & $ 1.3 \times 10^{-4}$ & $1.3 \times 10^{-3}$\\
Co        & $ 1.0 \times 10^{-3}$ & $ 1.0 \times 10^{-3}$ & $<1.0 \times 10^{-5}$ & $4.2 \times 10^{-6}$\\
Ni        & $ 7.1 \times 10^{-4}$ & $ 7.1 \times 10^{-4}$ & $ 5.0 \times 10^{-4}$ & $7.1 \times 10^{-5}$\\
\noalign{\smallskip} \hline
\end{tabular} 
\tablefoot{  \tablefoottext{a}{Abundances in mass fractions (see also Fig.\,\ref{fig:abundances}) and
    surface gravity $g$ in cm\,s$^{-2}$. }
  \tablefoottext{b}{Solar abundances from
    \citet{2009ARA&A..47..481A}.}  } 
\end{center}
\end{table}

\begin{table}[t]
\begin{center}
\caption{Number of non-LTE levels and lines of model ions.\tablefootmark{a} }
\label{tab:modelatoms} 
\tiny
\begin{tabular}{cccccccccc}
\hline 
\hline 
\noalign{\smallskip}
   & I       & II      & III      &  IV     &    V     &   VI    \\
\hline 
\noalign{\smallskip}
H  & 15, 105 \\
He & 29, 60  & 15, 105 \\
C  &         & 1, 0    & 133, 745 & 54, 279  \\   
N  &         &         & 1, 0     & 76, 405 & 54, 297  \\
O  &         &         &          & 83, 637 & 105, 671 & 54, 280 \\
Si &         &         & 17, 28   & 30, 102 & 25, 59   \\
P  &         &         &  3, 0    & 21, 9   & 18, 12   \\
S  &         &         &          & 39, 107 & 25, 48   & 38, 120 \\
\noalign{\smallskip} \hline
\end{tabular} 
\tablefoot{ \tablefoottext{a}{First and second number of each table
    entry denote the number of levels and lines, respectively. Not
    listed for each element is the highest considered ionization stage, which
    comprises its ground state only. See text for the treatment of
    iron-group elements.}  } 
\end{center}
\end{table}

\section{Ultraviolet observations}
\label{sect:observations}

We obtained HST/GHRS UV spectra of the three program stars
(Tab.\,\ref{tab:obs};
Figs.\,\ref{fig:hs2115_fuse}--\ref{fig:he0504_3000A.ps}). Without
quantitative analyses, these spectra were presented previously in
\cite{1997ASSL..214..207W,1997fbs..conf..227W,1999JCoAM.109...65W}. A
high-resolution segment of \henull\ at
2957--3000\,\AA\ (Fig.\,\ref{fig:he0504_3000A.ps}) was recorded along
with a red optical spectrum of the cool companion of
\hsnull\ (Fig.\,\ref{fig:hs0713_g750l_fit}).

Far-UV observations (915--1220\,\AA) of \henull\ and \hsnull\ were
taken with Berkeley Extreme and Far-UV Spectrometer (BEFS) aboard
ORFEUS-SPAS\,II in 1996, but the resolution (0.5\,\AA) was claimed to
be too poor for spectral line identification
\citep{1999ASPC..169..511W}. A reinspection of these data in the MAST
archive reveals that the spectrum of \hsnull\ is indeed very noisy,
while the \henull\ spectrum is better but still inferior to the
coadded FUSE data taken later. Hence we do not use the BEFS data. The
FUSE spectra of both stars were presented by
\cite{2003whdw.conf..171W} and the only photospheric lines identified
at that time were the \ion{S}{vi} and \ion{P}{v} resonance
doublets. Archival FUSE data of the DAO \hszwei\ and DO \hsx\ are used
for our work. They were hitherto unpublished and not analyzed. The
FUSE data were processed using the procedure described by
\cite{2015A&A...582A..94W}. All spectra were shifted such that the
photospheric lines have zero radial velocity. 

The FUSE observations exhibit many absorption lines of the
interstellar medium (ISM). To unambiguously identify stellar lines, we
employed the program OWENS
\citep{2002ApJS..140...67L,2002P&SS...50.1169H,2003ApJ...599..297H}. It
has the capacity to consider several, individual ISM clouds with their
own radial and turbulent velocity, temperature, chemical composition,
and respective column densities. We identified and modeled lines of
\ion{H}{i}, \ion{D}{i}, H$_2$ ($J$ = 0--9), HD ($J$ = 0--1),
\ion{C}{i--iii}, \ion{C}{ii}$^*$, \ion{N}{i--ii}, \ion{O}{i},
\ion{Si}{ii}, \ion{P}{ii}, \ion{S}{iii}, \ion{Ar}{i}, and
\ion{Fe}{ii}.

\section{Model atoms and model atmospheres}
\label{sect:models}

We used the T\"ubingen Model-Atmosphere Package
(TMAP\footnote{\url{http://astro.uni-tuebingen.de/~TMAP}}) to compute
non-LTE, plane-parallel, line-blanketed atmosphere models in radiative
and hydrostatic equilibrium
\citep{1999JCoAM.109...65W,2003ASPC..288...31W,tmap2012}.
Table\,\ref{tab:modelatoms} summarizes the number of considered
non-LTE levels and radiative transitions between them. All model atoms
were built from the publicly available T\"ubingen Model Atom Database
\citep[TMAD;][]{2003ASPC..288..103R}, which is comprised of data from
various sources, namely \citet{1975aelg.book.....B}, the databases
of the National Institute of Standards and Technology
(NIST\footnote{\url{http://www.nist.gov/pml/data/asd.cfm}}), the Opacity Project
\citep[OP\footnote{\url{http://cdsweb.u-strasbg.fr/topbase/topbase.html}};][]{1994MNRAS.266..805S},
CHIANTI\footnote{\url{http://www.chiantidatabase.org}}
\citep{1997A&AS..125..149D,2013ApJ...763...86L}, and the
Kentucky Atomic Line
List\footnote{\url{http://www.pa.uky.edu/~peter/atomic}}. 

Besides H, He, C, N, and O, our models include Si, P, S, Cr, Mn, Fe,
Co, and Ni. To reduce the computational efforts, the light metals (up
to sulfur) were considered with small model atoms and subsequently,
one by one, dealt with large model atoms while keeping fixed the
atmospheric structure. The statistics of these large model atoms
is summarized in Tab.\,\ref{tab:modelatoms}. 

For the iron-group elements considered (Cr--Ni), we used a statistical
approach, employing seven superlevels per ion linked by superlines,
together with an opacity sampling method
\citep{1989ApJ...339..558A,2003ASPC..288..103R}.  Ionization stages
\ion{}{iv--viii} augmented by single, ground-level stages \ion{}{ix}
were considered. We used the complete line list of Kurucz
\citep[so-called LIN lists, comprising about $4.5\times10^5$,
  $1.2\times10^6$, $7.6\times10^6$, $3.5\times10^6$, and
  $1.8\times10^6$ lines of the considered ions of Cr, Mn, Fe, Co, and
  Ni;][]{kurucz1991,kurucz2009, kurucz2011} for the computation of the
non-LTE population numbers, and the so-called POS lists, which include
only the subset of lines with well-known, experimentally observed line
positions, for the final spectrum synthesis.

\section{Spectral line fitting procedure and results}
\label{sect:results}

\begin{figure}[t]
 \centering  \includegraphics[width=1.0\columnwidth]{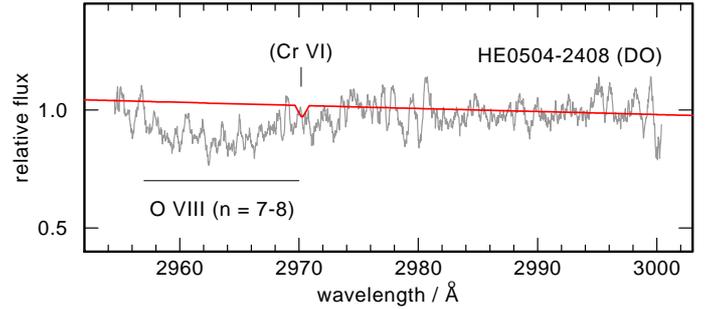}
  \caption{HST G270M spectrum of \henull\ compared to the final
    model. The broad absorption feature is a \emph{uhe} line as
    labeled.
}\label{fig:he0504_3000A.ps}
\end{figure}

For the line profile fitting we proceeded as follows. In the
1150--1300\,\AA\ range of the HST spectra we assumed an arbitrary
normalization factor and a reddening of $E(B-V)=0.13$ to match the
continuum shape. In all other wavelength windows of the HST and FUSE
spectra shown in
Figs.\,\ref{fig:hs2115_fuse}--\ref{fig:hs0713_g750l_fit}, zero
reddening was assumed and a flux scaling factor that can be different
from window to window was applied. All model spectra were convolved
with Gaussians according to the resolution of the observations.

The observed flux falls significantly below the models at
about 1490--1520\,\AA\ in the spectra of \henull\ and \hsnull,\ which we
assign to the blueshifted \ion{O}{vii} n=5$\rightarrow$6 \emph{uhe}
line (hydrogenic wavelength 1522\,\AA). Another broad feature at
1390-1430\,\AA\ remains unexplained. In the G270M spectrum of
\henull\ (Fig.\,\ref{fig:he0504_3000A.ps}) we identify the blueshifted
\ion{O}{viii} n=7$\rightarrow$8 \emph{uhe} line covering the
2957--2970\,\AA\ interval (hydrogenic wavelength 2979\,\AA). This line is so
broad that it is even detectable in a low-resolution (about 5\,\AA)
IUE\footnote{International Ultraviolet Explorer} spectrum (image
LWP22367, taken from the MAST archive). The feature is also detectable
in a IUE spectrum of \hsnull\ (LWP27020). In the same IUE spectra, the
blueshifted \ion{O}{vii} n=6$\rightarrow$7 \emph{uhe} line (hydrogenic
wavelength 2525\,\AA) is present in both DOs, and at shorter
wavelengths the blueshifted \ion{O}{viii} n=6$\rightarrow$7 \emph{uhe}
line (hydrogenic wavelength 1933\,\AA) is just detectable (images
SWP43954 and SWP49562, respectively).

\subsection{\hszwei\ (DAO)}

\subsubsection{Effective temperature and surface gravity}

At the outset, we computed a grid of hydrogen models with a small
admixture of helium (He = 0.004; all element abundances in this paper
are given in mass fractions) to constrain \logg\ and \Teff\ from the
Lyman lines in the FUSE spectrum. The value of \Teff\ was further
constrained with a series of models including metals by exploiting
several ionization balances (\ion{C}{iii/iv}, \ion{N}{iv/v},
\ion{O}{v/vi}, \ion{Fe}{v/vi}, \ion{Ni}{v/vi}). We finally adopted
\Teff = $80\,000\pm5000$\,K and \logg = $7.0\pm0.5$. While the
uncertainty in \Teff\ is rather small, a tighter limit to \logg\ is
not possible. The significantly lower values for \Teff\ found in
previous analyses of optical spectra (67\,000\,K and 62\,230\,K; see
Sect.\,\ref{sect:dao}) are clearly excluded by our model fits to the
metal lines, and the high surface gravity derived by
\citet{2010ApJ...720..581G} from the Balmer lines (\logg = 7.76)
contradicts the Lyman-line wings. In their original discovery paper,
\citet{1995A&A...303L..53D} derived \logg = 6.9 from Balmer-line
fitting with pure H/He NLTE models, which is close to our value; these
authors note, however, the fit is poor and has a pronounced Balmer
line problem.

\subsubsection{Element abundances}

\paragraph{Helium.} \ion{He}{ii}\,$\lambda$1640\,\AA\ suggests He = 0.001, which is lower
than derived previously from \ion{He}{ii}
$\lambda$4686\,\AA\  \citep[0.006 and 0.004;
][]{1995A&A...303L..53D,2010ApJ...720..581G}. Performing a fit to
\ion{He}{ii}\,$\lambda$4686\,\AA\ in the spectrum presented by
\cite{1995A&A...303L..53D} with our models confirms their high value
of He = 0.006, i.e., we obtain no consistent fit to the UV and optical
\ion{He}{ii} lines. At any rate, because of the low abundance, no
\ion{He}{ii} lines are detectable in the FUSE spectrum.

\paragraph{Light metals.} From a few \ion{C}{iv} lines we find C =
$3.0 \times 10^{-4}$. The resonance doublet is not deep enough in the
model and we suspect an interstellar contribution to the observed
profile. The \ion{C}{iii} multiplet at 1175\,\AA\ is not detectable,
confirming the high \Teff. From a few \ion{N}{iv} lines including
$\lambda$1718\,\AA\ and the \ion{N}{v}
$\lambda\lambda$1239/1243\,\AA\ resonance doublet we conclude N = $3.0
\times 10^{-6}$. \ion{O}{v}~$\lambda$1371\,\AA\ and the \ion{O}{vi}
$\lambda\lambda$1032/1038\,\AA\ resonance doublet suggest O = $ 7.0
\times 10^{-5}$. The \ion{O}{iv} triplet at 1339/1343/1344\,\AA\ is
not detectable, again confirming the high temperature. From the
\ion{Si}{iv} $\lambda\lambda$1394/1403\,\AA\ resonance doublet and the
\ion{Si}{iv} $\lambda\lambda$1122/1128\,\AA\ doublet, we find Si = $
3.0 \times 10^{-5}$. From the resonance doublets \ion{P}{v}
$\lambda\lambda$1118/1128\,\AA\ and \ion{S}{vi}
$\lambda\lambda$933/944\,\AA, P = $ 1.0 \times 10^{-6}$ and S = $ 5.0
\times 10^{-6}$ is derived.

\paragraph{Iron-group elements.}

The UV spectra are dominated by lines from iron and nickel. From iron
we see mainly \ion{Fe}{v} lines, which are predominantly located in
the 1360--1430\,\AA\ band. In comparison, only few \ion{Fe}{vi} lines
can be identified. We derive a solar abundance (Fe = $ 1.3 \times
10^{-3}$). From nickel, we see lines from \ion{Ni}{v} (most of them in
the 1230--1350\,\AA\ region) and \ion{Ni}{vi} (1070--1200\,\AA) and we
adopt a 10 times solar abundance (Ni = $ 7.1 \times 10^{-4}$).

As indicated in Figs.\,\ref{fig:hs2115_fuse} and \ref{fig:hs2115_hst},
we identify lines from chromium, manganese, and cobalt. Specifically,
we detect a smaller number of \ion{Cr}{v} and \ion{Cr}{vi} lines and
derive Cr = $ 1.3 \times 10^{-3}$, which is about 2 dex oversolar. A
few \ion{Mn}{vi} lines are detectable and we measure a slightly
oversolar abundance (Mn = $ 3.0 \times 10^{-5}$). A larger number of
\ion{Co}{vi} lines can be seen, contributing to the nickel line forest
in the 1120--1260\,\AA\ range. The derived abundance of Co = $ 1.0
\times 10^{-3}$ is more than two dex oversolar.  The error in the
abundance determinations is estimated to $\pm$0.3~dex.

We comment on the many unidentified lines in this and the other
program stars in Sect.\,\ref{sect:unid} below.

\begin{figure}[t]
 \centering
 \includegraphics[width=1.0\columnwidth]{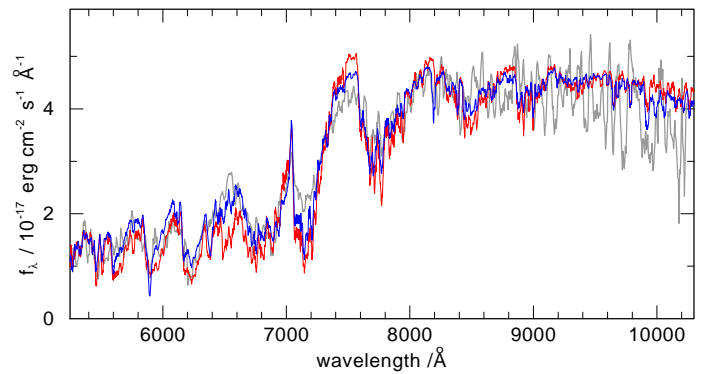}
  \caption{HST/STIS G750L spectrum of the cool companion of
    \hsnull\ (gray line) compared to two NextGen models (blue: \Teff =
    3300\,K, \logg = 5; red: \Teff = 3600\,K; \logg = 2.5). Models and
    observation were smoothed with a 10\,\AA\ wide boxcar.
  }\label{fig:hs0713_g750l_fit}
\end{figure}

\begin{figure*}[t]
 \centering  \includegraphics[width=1.0\textwidth]{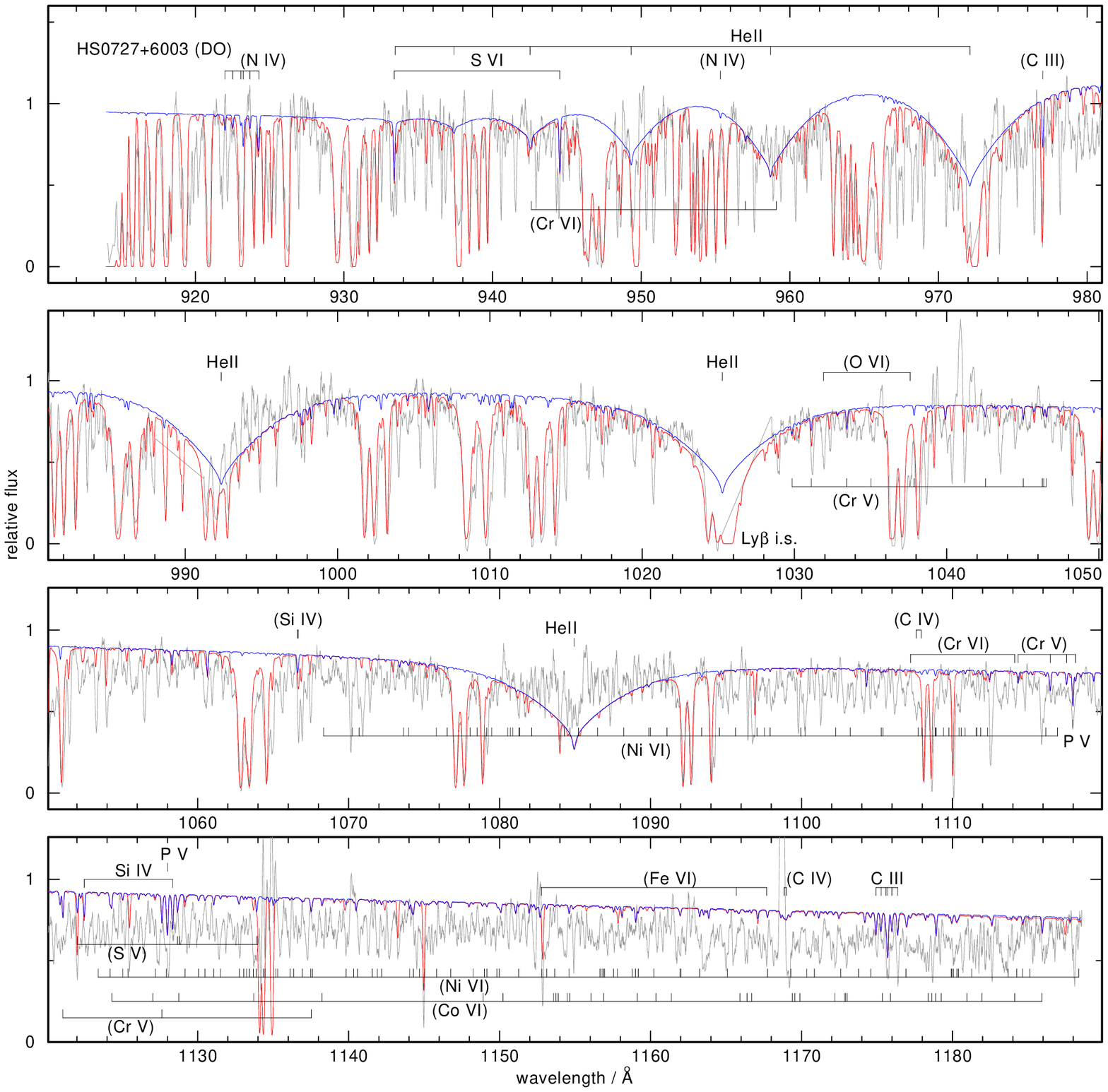}
  \caption{FUSE spectrum of the DO \hsx, similar to
    Fig.\,\ref{fig:hs2115_fuse}. Model: \Teff = 65\,000\,K, \logg = 7.5
    (same one as for \hsnull\ in Fig.\,\ref{fig:hs0713_fuse}).}
\label{fig:hs0727_fuse}
\end{figure*}

\subsection{\henull\ (DO)}

\subsubsection{Effective temperature and surface gravity}

In analogy to the DAO analysis described above, we began our analysis
of both DO stars with a grid of pure helium models (with a small
admixture of $2.5\times 10^{-4}$ hydrogen) to roughly constrain
\Teff\ and \logg, and then fine-tuned \Teff\ by exploiting metal
ionization balances. For \henull\ we found \Teff =
$85\,000\pm10\,000$\,K and \logg = $7.0\pm0.5$, i.e., similar values
as for the DAO analysis, but the uncertainty in \Teff\ is twice as large,
mainly because of the weaker lines of C, N, and O. A lower temperature
is excluded by the lack of the \ion{C}{iii} multiplet at
1175\,\AA\ and the lack of \ion{O}{iv} lines in the HST
spectrum. Fittingly, in a high-resolution (0.1\,\AA) optical spectrum
\citep[taken in the SPY survey; ][]{2003Msngr.112...25N} the
\ion{He}{i} 5876\,\AA\ line is not detectable, yielding a lower limit
of \Teff = 80\,000\,K. Higher temperatures are excluded because
low-ionization heavy-metal lines would disappear (\ion{Cr}{v},
\ion{Fe}{v}, \ion{Ni}{v}). Higher surface gravities would give a too
broad \ion{He}{ii} $\lambda$1640\,\AA\ line, while lower gravities
would give too deep, narrow line cores in the higher series members of
the Pickering series in the optical.

The model poorly fits the \ion{He}{ii} lines at 991\,\AA\ and
1085\,\AA, which appear much shallower in the observation. A possible
explanation is discussed in Sect.\,\ref{sect:unid}.

\subsubsection{Element abundances}

We derive C, N, and O abundances that are lower than in the DAO star
\hszwei. For N, only an upper limit could be fixed (C = $3.0\times
10^{-5}$, N $<1.0 \times 10^{-6}$, O = $ 1.0 \times 10^{-5}$). While
the Si abundance is the same ($ 3.0 \times 10^{-5}$), significantly
larger P and S abundances were found (P = $ 1.0 \times 10^{-5}$, S =
$ 5.0 \times 10^{-5}$). For Cr, Mn, Fe, Co, and Ni we derive the same
abundances as in the DAO analysis. As before, the error in the abundance
determinations is estimated to $\pm$0.3~dex.

\subsection{\hsnull\ (DO)}

\subsubsection{Effective temperature and surface gravity}

As mentioned in Sect.\,\ref{sect:dos}, a weak \ion{He}{i}
5876\,\AA\ line is detectable in the SDSS spectrum and serves as a
good constraint for \Teff. At \logg = 7.5 we find a best fit at \Teff
= $65\,000 \pm 5000$\,K. Below this temperature, \ion{He}{i}
4471\,\AA\ would be detectable but it is not visible in the
observation. \citet{2014A&A...566A.116R} derived a higher temperature
(\Teff = 80\,000$\pm$10\,000\,K and \logg = 7.75$\pm$0.5) using
helium-model atmospheres including only carbon. Our pure helium-models
confirm this result. But in our fully metal line blanketed models, the
\ion{He}{i} 5876\,\AA\ line is significantly weaker so that we arrive
at a lower temperature for a good fit. The surface gravity cannot be
fixed better from the optical spectra. For the following we assume
\logg = 7.5$\pm$0.5. The error range comprises the previously
estimated values for the gravity. 

As was the case with the other DO, the model poorly fits the
\ion{He}{ii} lines at 991\,\AA\ and 1085\,\AA\ because these lines are
too deep to fit the observation (see discussion in
Sect.\,\ref{sect:unid}).

\subsubsection{Element abundances}\label{sect:hsnullabu}

\paragraph{Light metals.} The carbon abundance is solely derived from the 
\ion{C}{iii} multiplet at 1175\,\AA\ (C = $ 7.0 \times 10^{-6}$). From
the  absence of \ion{N}{iv} and \ion{N}{v} lines and the
\ion{O}{iv} multiplet at 1338.61/1342.99/1343.51\,\AA\ and
\ion{O}{v}~1371\,\AA, we derive N $< 3.0 \times 10^{-7}$ and O $< 3.0
\times 10^{-6}$. The \ion{O}{vi} resonance doublet is absent because
of the low \Teff. For the other light metals we find from the same
lines as discussed above Si = $5.0 \times 10^{-6}$, P = $1.5 \times
10^{-7}$, and S = $1.0 \times 10^{-5}$. Quite large errors of 0.7
dex must be accepted because of the relatively poor quality of the
spectra.

\paragraph{Iron-group elements.} In the models, lines of \ion{Cr}{v} are 
dominant while \ion{Cr}{vi} lines are weak. However, none of the lines
can be identified in the observation and we find Cr $< 1.3 \times
10^{-4}$. The \ion{Mn}{vi} lines seen in the two other stars are
absent here because of the lower temperature. No other Mn lines are
predicted so that we are unable to derive an upper abundance limit. As to
iron, \ion{Fe}{v} lines are stronger than \ion{Fe}{vi} lines in the
models. We find Fe = $ 1.3 \times 10^{-4}$. Lines from cobalt cannot
be identified. Mainly from the absence of \ion{Co}{v} lines in the HST
range, we fix Co $< 1.0 \times 10^{-5}$. From \ion{Ni}{v} lines we
measure Ni = $ 5.0 \times 10^{-4}$. Again, relatively large errors of
0.7 dex must be accepted.

\paragraph{Trans-iron elements.} In the FUSE spectrum of \hsnull, we
possibly found lines from heavier elements but the spectrum is too
noisy for firm detections. By comparison with the DO white dwarf
RE\,0503$-$289, which has similar \Teff\ and \logg\ (75\,000\,K, 7.5)
and in which 14 trans-Fe elements were discovered
\citep{2017A&A...598A.135H}, lines from
\ion{Zn}{v}, \ion{Ge}{v/vi}, \ion{Se}{vi}, \ion{Mo}{vi}, \ion{Sn}{v},
and \ion{Te}{vi} might be present.

\begin{figure*}[t]
 \centering  \includegraphics[width=1.0\textwidth]{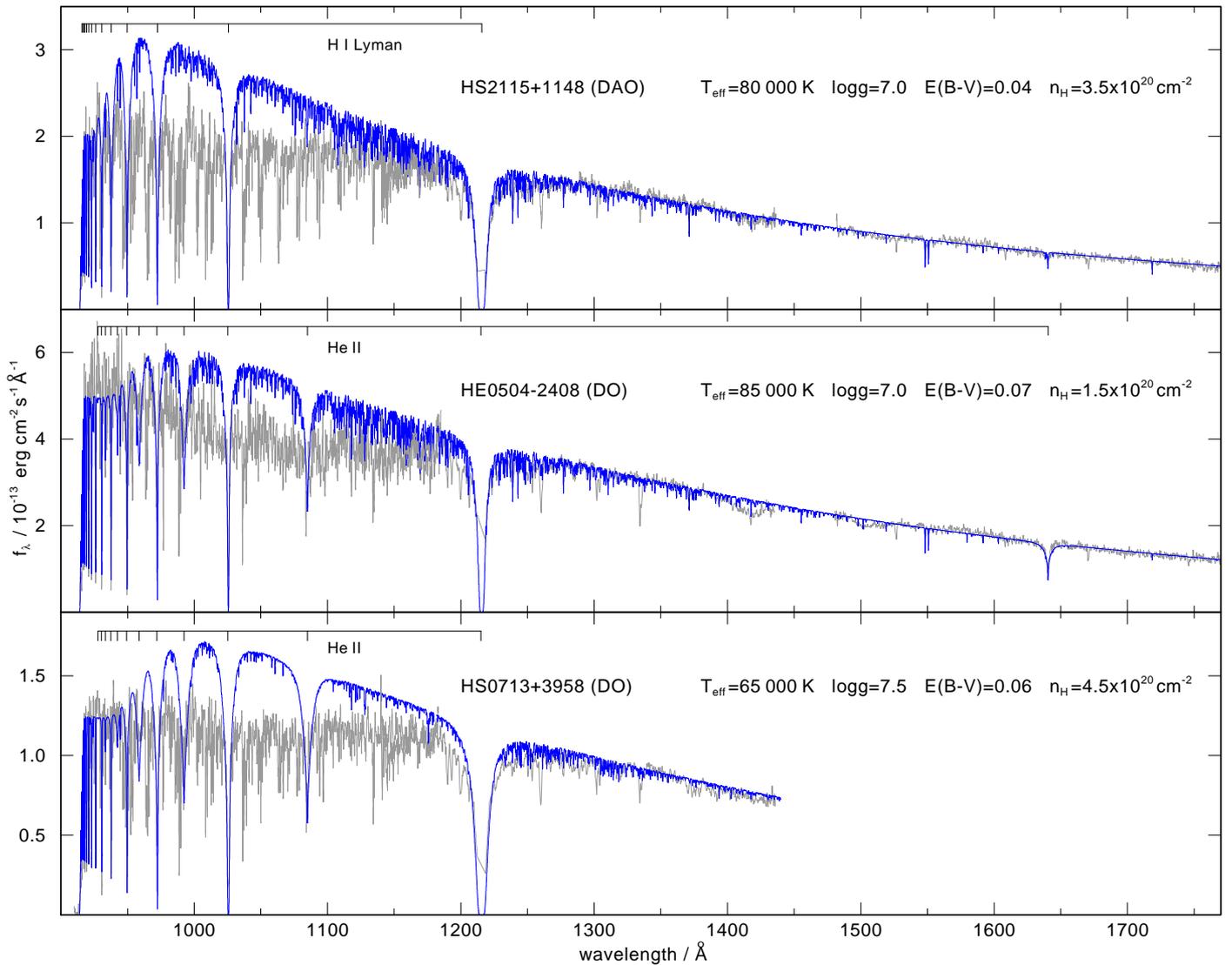}
  \caption{Absolute flux distributions of our program stars (gray
    lines) compared to the final photospheric models (blue lines). The
    models include absorption by interstellar hydrogen Lyman lines and
    reddening according to the column densities $n_{\rm H}$ and
    $E(B-V)$ values as given in the panels. For clarity, models and
    observations were smoothed with Gaussians with a full width at
    half maximum of 0.3\,\AA.}
\label{fig:overview}
\end{figure*}

\subsubsection{Spectroscopy of the cool companion} 

In Fig.\,\ref{fig:hs0713_g750l_fit}, we show the HST/STIS G750L
spectrum that we took from the cool companion of \hsnull. It is
characterized by TiO bands. We compare it to two NextGen
\citep{2001ApJ...556..357A} models (solar abundances) that suggest
that \Teff = 3300\,K and \logg = 5, hence, the spectral type is M5V
\citep[][p. 312; citing \citet{1982lbg6.conf.....A}]{voigt...abriss},
which is later than estimated previously from photometry. 

The quality of the spectrum is not good enough to discern between
models with different surface gravity (i.e., luminosity). A similarly
good fit is achieved with a subgiant spectrum with \Teff = 3600\,K and
\logg = 2.5, however, the companion must be a dwarf. The spectroscopic
distance to the WD primary is 392\,pc, derived from the ratio of the
observed to the model flux, the surface gravity \logg = 7.5$\pm$0.5,
and the spectroscopic mass of M$_{\rm WD}$ = 0.45$\pm$0.06\,M$_\odot$
\citep[from extrapolation of evolutionary tracks
  from][]{2009ApJ...704.1605A}. The uncertainty in the distance is
about a factor of two, which is dominated by the error in surface
gravity. Taking this distance, we derive a radius of the companion of
R$_{\rm comp}=0.4$\,R$_\odot$, again within a factor of two, by
comparing the observed and model (\Teff = 3300\,K, \logg = 5)
spectra. It is consistent with the radius of a M5V star
\citep[0.27\,R$_\odot$;][p. 324]{voigt...abriss}.

\subsection{Archival FUSE data of \hsx\ (DO)}
\label{sect:hsx}

As mentioned in the Introduction, we retrieved a previously
unpublished FUSE spectrum of yet another \emph{uhe} DO, \hsx, from the
MAST archive. It is shown in Fig.\,\ref{fig:hs0727_fuse}, together
with the final model spectrum for \hsnull,\ which we have shown in
Fig.\,\ref{fig:hs0713_fuse}. The presence of the \ion{C}{iii}
multiplet at 1175\,\AA\ indicates a similar temperature as that of
\hsnull\ (65\,000\,K) but without further UV data, no detailed
assessment of the photospheric parameters is possible. The only other
metal lines that can be identified are from Si, P, and S, and their
strengths point at similar abundances as in \hsnull. 

Two things are remarkable. First, even more discrepant than in the two
analyzed DOs is the very weak \ion{He}{ii} 1085\,\AA\ line, which is
much deeper in the model. Second, we are unable to identify any line
of iron-group elements (see discussion in Sect.\,\ref{sect:unid}).

\subsection{Unidentified opacities in the program stars}\label{sect:unid}

In the spectra of all three program stars, particularly in the higher
resolved FUSE data, many spectral lines remain unidentified. In fact,
if we look at the overall flux distributions, we see that extreme
blanketing by unidentified lines causes significant flux depression in
the 950--1150\,\AA\ spectral range
(Fig.\,\ref{fig:overview}). According to our models, they cannot be
assigned to iron-group elements or other light metals. Instead, we
might see either a multitude of lines from trans-iron elements (some
of which are possibly identified in \hsnull; see
Sect.\,\ref{sect:hsnullabu}) and/or the flux depressions could
stem from broad, unknown opacity sources, for example, from superionized
species.

We ruled out that calibration problems cause the 950--1150\,\AA\ flux
depression. The fundamental FUSE flux calibration is based on the same
DA WDs as the HST spectrographs, so there is no disconnect in the
definition that might lead to spurious spectral slope
determinations. Furthermore, the deviation in spectral slope between
the models and data begins around Ly\,$\alpha$ in the two hotter
stars and longward of Ly\,$\alpha$ in the cooler \hsnull, well within
the STIS band. The four FUSE channels are each calibrated separately
and the measured spectra in the four channels were consistent within a
few percent, which is consistent with the scatter among the flux calibration
measurements obtained throughout the FUSE mission. Similarly,
measurements of the program stars obtained at multiple epochs were
also consistent to within a few percent, indicating that there were no
shortcomings in the corrections for the gradually varying throughput
of the instrument. (See Section 7.5 of the FUSE Data Handbook for
information on the FUSE flux calibration and performance throughout
the mission\footnote{\url{http://archive.stsci.edu/fuse/dh.html}}.)

The slope of the FUV extinction curve is known to vary from one
Galactic line of sight to another, but all derived reddening laws have
been monotonic shortward of about
1800\,\AA\ \citep[e.g.,][]{2005ApJ...625..167S,2005ApJ...630..355C}. There
has been no evidence for a bump in the FUV extinction curve at
1050\,\AA\ analogous to the well-known bump at 2175\,\AA.

Fig.\,\ref{fig:overview} also explains the weakness of the
\ion{He}{ii} lines in the FUSE spectra of the DOs. In the previous
figures, where we have shown the FUSE spectra, we normalized
the model spectra to the local continua in the panels, however, it is
obvious that the true continua are undetermined because of strong line
blanketing. This finding calls into question the validity of our
abundance determinations of metals, which exclusively rely on lines
located in the range 950--1150\,\AA. This mainly affects phosphorus
and sulfur, whose abundances could be systematically higher than we
derived.

As mentioned previously (Sect.\,\ref{sect:results}), there is a very
shallow, roughly 20\,\AA\ wide absorption feature in \henull\ centered
at around 1500\,\AA. As an alternative identification as a possible
\emph{uhe} line, we speculated that it might be the blueshifted
component of a weak \ion{C}{iv}\,$\lambda$1550\,\AA\ P~Cygni profile
\citep{2003whdw.conf..171W}. If true, then it would stem from a wind
with a terminal velocity of about 10\,000\,km/s.

\subsection{Optical spectra}

We compared our final models for the three program stars with
optical spectra. In all cases, the problems with the too-deep hydrogen
Balmer and \ion{He}{ii} lines remained. The problems of our models
could be related to the lack of opacities from unidentified
species. In the case of the DAO, the H$\delta$ and H$\epsilon$ lines
fit well (higher series members are not present) while the lower
series members are still too weak in the model, as in previous
work. In the two DOs, the same problem holds for the \ion{He}{ii}
Pickering lines and \ion{He}{ii} $\lambda$4686\,\AA. The relatively
low \logg = 7 in the models is not in contradiction to the
non-detection of Pickering lines shortward of \ion{He}{ii}
$\lambda$4340\,\AA. We reiterate that in the case of the DAO, we
  obtain inconsistent He abundances derived from the \ion{He}{ii}
  lines at $\lambda$1640\,\AA\ and $\lambda$4686\,\AA, namely, He =
  0.001 and 0.006.

\begin{figure}[t]
 \centering  \includegraphics[width=1.0\columnwidth]{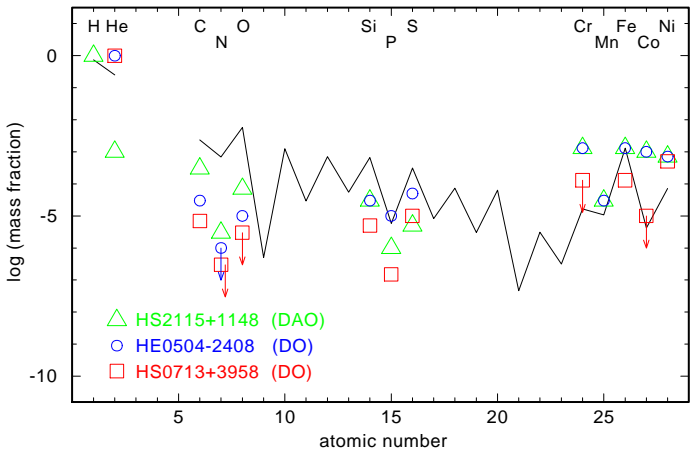}
  \caption{Abundances measured in the three program stars. The black
    line indicates solar values.}\label{fig:abundances}
\end{figure}

\section{Summary and discussion}
\label{sect:discussion}

We determined metal abundances in three hot white dwarfs showing
signatures of a superionized wind (see Table~\ref{tab:stars} and
Fig.\,\ref{fig:abundances}). Generally, we find low abundances of the
light elements C, N, O, Si, P, and S (between solar and less than
$10^{-3}$ solar), while iron is solar and other Fe-group elements
are between roughly solar and 100 times solar. The abundance patterns are
not unusual when compared to other hot WDs with similar effective
temperature. It can be assumed that they are determined by
gravitational settling and radiative acceleration, although
\citet{2014MNRAS.440.1607B} argued that the wide spread of observed
element abundances in hot DA white dwarfs in a given temperature range could
result from accretion from external sources. 

In particular, the DAO \hszwei\ (\Teff = 80\,000\,K, \logg = 7) has a
metal abundance pattern that is very similar to BD$-22^\circ 3467$
\citep[80\,000\,K, 7.2;][]{2012A&A...548A.109Z} with respect to all
species. Because of the
nearly identical photospheric parameters one might ask whether
BD$-22^\circ 3467$ exhibits \emph{uhe} lines in the optical
spectrum. Unfortunately, no optical spectrum exists because it is a
close binary star with a dominating cool companion.
\citet{2005MNRAS.363..183G} measured abundances of C, N, O,
Si, Fe, and Ni in 16 DAO white dwarfs and the results are
qualitatively similar to those for \hszwei\ as well. 

We seek to understand how the metal abundances of the two DOs of our
investigation (having 65\,000\,K and 85\,000\,K; both with \logg = 7)
compare to other DOs. A recent investigation of two DOs with \logg=7
but significantly higher temperatures (115\,000\,K and 125\,000\,K)
revealed solar abundances of heavier elements (Ne, Si, P, S, Ar, Fe,
and Ni). This is the consequence of on-going mass loss preventing
gravitational settling \citep{2017A&A...601A...8W}. The most detailed
analyzed DO with similar temperature is \re\ \citep[70\,000\,K, \logg
  = 7.5; ][]{2017rauch}. With the exception of carbon, which is
oversolar in \re, other light metals (N, O, Si, P, and S) are
underabundant by about one dex, in qualitative agreement with our DOs
studied here. Fe and Ni abundances in \re\ ($<0.01$ solar and solar,
respectively) are lower than in our two DOs (0.1--1.0 solar and 0.9--1
solar). Aside from the enhanced Fe and Ni abundances, there is no
significant difference in the abundances of the two hot-wind DOs and
\re, which has no signatures of a superhot wind.

For our analysis, we computed line-blanketed, chemically
homogeneous non-LTE model atmospheres, including the opacities of all
investigated metals. We hoped that these models would remove the
Balmer-line problem in the DAO and the respective \ion{He}{ii} line
problem in the DOs. However, the effect of the considered metal
opacities on the atmospheric structure is insufficient to obtain good
fits to the extraordinarily deep H and \ion{He}{ii} line cores in the
optical spectra. This corroborates our earlier suspicion that this
problem is probably related to the occurrence of the
ultra-high ionization lines associated with a super-hot wind.

In the UV spectra, there is no hint as to ongoing mass loss. There is
neither blueshift of nor asymmetry in the resonance line profiles
(doublets of \ion{C}{iv}, \ion{N}{v}, \ion{O}{vi}, \ion{Si}{iv},
\ion{P}{v}, \ion{S}{vi}) or other features like
\ion{O}{v}~$\lambda$1371\,\AA\ and \ion{He}{ii}~$\lambda$1640\,\AA
\ (for a possible exception concerning the \ion{C}{iv} line, see
Sect.\,\ref{sect:unid}). This is not unexpected because mass-loss
rates predicted from radiation-driven wind theory are so low
\citep[\md $<10^{-13}$;][]{2000A&A...359.1042U} that they cannot be
detected spectroscopically in these lines. On the other hand, if we
believe that the \emph{uhe} line features are signatures of a
superionized wind, then the question arises where these features are
formed and how the wind is heated to temperatures on the order
$10^6$\,K. The situation is reminiscent of the occurrence of
`superionization' in the winds of extreme helium (EHe) stars
\citep{1982A&A...116..273H}, albeit at a lower temperature scale. For
example, \citet{2010MNRAS.404.1698J} in their investigation of  mass
loss in the very thin wind of \bd\ (\Teff= 18\,500\,K, \logg = 2.6),
the observed wind profile of the \ion{C}{iv} resonance line could not
be reproduced by the models because at that low \Teff\ they predicted
that carbon recombines to \ion{C}{iii} and \ion{C}{ii}. It was
concluded that some `superionization' keeps the ionization higher than
predicted. As a viable mechanism these authors discussed frictional
heating in a multicomponent wind. In fact, stellar wind models for
subluminous hot stars confirm that this is well possible
\citep{2006BaltA..15..147U,2016A&A...593A.101K}. In the present
context of the much hotter WDs, such `superionization' is observed,
albeit involving higher temperatures and, thus, higher ionization
stages.

It has been predicted by models that the process of frictional heating
can occur in thin radiation-driven winds of hot stars
\citep{1992A&A...262..515S}. Radiation pressure accelerates only metal
ions while the bulk matter (hydrogen and/or helium) stays inert. In
models for such multicomponent wind models for OB stars, temperatures
on the order $10^6$\,K were predicted
\citep{2001A&A...377..175K}. Particularly interesting for our
discussion here are investigations of stellar winds with low
metallicity because in the WDs of the present paper, the abundances of
light metals are well subsolar. These investigations aimed at stars
with low metallicity in the early Universe or even at the first stars
which, in late evolutionary phases, have CNO driven winds
\citep{2010A&A...516A.100K}. For example, for a massive star with
\Teff = 40\,000\,K and 0.001 solar metallicity,
\citet{2003A&A...402..713K} find a strong increase in temperature of
the photon absorbing ions to T $\approx$ 200\,000\,K close to the
stellar surface (1.14 stellar radii), while the non-absorbing bulk
component (hydrogen and helium) remains at temperatures below
$100\,000$\,K. It is predicted, that the bulk material either forms
clouds around the star or falls back to the stellar surface. We may
speculate that the too-deep optical H and \ion{He}{ii} lines in our
WDs are caused by relatively cool gas in such a circumstellar static
cloud, while the \emph{uhe} lines form in the same region but
originate from the frictionally heated metal ions, which were
accelerated to develop a superhot high-velocity wind. A possible
scenario of pure metallic winds with hydrostatic H and He at \md $ <
10^{-16}$ has been discussed in the context of subdwarf B stars
\citep{2008A&A...486..923U} and this could be relevant in the
superionized wind WDs as well.

Systematic studies of multicomponent wind models for hot white dwarfs
are necessary to check the possibility of whether the hot-wind phenomenon
is indeed caused by frictional heating in radiatively driven winds. In
a first study presented by \citet{2005ASPC..334..337K} it can be seen,
that a cooling white dwarf crosses a narrow strip in the
\logg--\Teff\ diagram where frictionally heated winds occur. The strip
edge at low temperature is defined by the complete decoupling of the
metallic wind from the bulk H+He matter, while the strip edge at high
temperature results from the transition to chemically homogeneous
winds because of high mass-loss rates. This strip could explain the
fact that the hot-wind WD phenomenon is restricted to WDs in the range
\Teff = 65\,000--120\,000\,K. Such a systematic study could also answer
the question why this phenomenon predominantly affects
helium-dominated WDs, while only one affected (He/C/O-dominated)
PG1159 star and one H-rich WD are known.

We close by mentioning that the first example of a star with a
fractionated wind is the Bp star $\sigma$~Ori~E
\citep{1997A&A...319..250G}. This star is an oblique magnetic rotator whose
field supports two corotating circumstellar clouds, fed by the
photospheric wind, and located at the intersections of the magnetic and
rotational equators. The gas temperature in the clouds is similar to
the stellar effective temperature and the clouds are opaque in the
lines and continuum \citep{2007A&A...475.1027S}, as derived from
variations of high-level Balmer lines first discussed by
\citet{1976A&A....52..303G,1977A&A....56..129G}. The gas further
expands into an outer, rotational-phase independent, coronal wind and,
according to \cite{1984A&A...138..421H}, could attain temperatures of
$10^5-10^7$\,K. We may speculate that such a rigidly rotating
magnetosphere model with an ambient, very hot coronal wind might be
related to the \emph{uhe} lines in the WDs discussed in the present
paper.

$\sigma$~Ori~E is variable, both spectroscopically and
photometrically. The period of about 1.19 days
\citep{1976Natur.262..116H,1977A&AS...30...11P} today is explained to
be the rotation velocity of the star and light variations are
considered as due to star spots \citep[see the most recent analysis
  of][]{2015MNRAS.451.2015O}. In fact, it is intriguing that the
Catalina Sky Survey found the \emph{uhe} DO \hsx\ (discussed in
Sect.\,\ref{sect:hsx}) to be variable in light with a period of 0.284
days and a photometric amplitude of 0.14~mag
\citep{2014ApJS..213....9D}. This short period could well be the
rotation period implying that the star may be spotted as well. In
fact, Reindl et al. (in prep.) reported both spectral and photometric
variability of \emph{uhe} DOs.

\begin{acknowledgements} 
We thank R\@. Napiwotzki for putting the SPY survey spectra at our
disposal. T\@. Rauch was supported by the German Aerospace Center
(DLR) under grant 50\,OR\,1507. The TMAD service
(\url{http://astro.uni-tuebingen.de/~TMAD}) used to compile atomic
data for this paper and the TIRO service
(\url{http://astro.uni-tuebingen.de/~TIRO}), used to generate the model
atoms for the iron-group elements, were constructed as part of the
activities of the German Astrophysical Virtual Observatory. This
research has made use of the SIMBAD database, operated at CDS,
Strasbourg, France, and of NASA's Astrophysics Data System
Bibliographic Services. Some of the data presented in this paper were
obtained from the Mikulski Archive for Space Telescopes (MAST). This
work had been carried out using the profile fitting procedure OWENS
developed by M\@. Lemoine and the FUSE French Team.
\end{acknowledgements}

\bibliographystyle{aa}
\bibliography{aa}

\begin{thebibliography}{64}
\expandafter\ifx\csname natexlab\endcsname\relax\def\natexlab#1{#1}\fi

\bibitem[{{Allard} {et~al.}(2001){Allard}, {Hauschildt}, {Alexander},
  {Tamanai}, \& {Schweitzer}}]{2001ApJ...556..357A}
{Allard}, F., {Hauschildt}, P.~H., {Alexander}, D.~R., {Tamanai}, A., \&
  {Schweitzer}, A. 2001, \apj, 556, 357

\bibitem[{{Aller} {et~al.}(1982){Aller}, {Appenzeller}, {Baschek}, {Duerbeck},
  {Herczeg}, {Lamla}, {Meyer-Hofmeister}, {Schmidt-Kaler}, {Scholz},
  {Seggewiss}, {Seitter}, \& {Weidemann}}]{1982lbg6.conf.....A}
{Aller}, L.~H., {Appenzeller}, I., {Baschek}, B., {et~al.}, eds. 1982,
  {Landolt-B{\"o}rnstein: Numerical Data and Functional Relationships in
  Science and Technology - New Series `` Gruppe/Group 6 Astronomy and
  Astrophysics '' Volume 2 Schaifers/Voigt: Astronomy and Astrophysics /
  Astronomie und Astrophysik `` Stars and Star Clusters / Sterne und
  Sternhaufen}

\bibitem[{{Althaus} {et~al.}(2009){Althaus}, {Panei}, {Miller Bertolami},
  {Garc{\'{\i}}a-Berro}, {C{\'o}rsico}, {Romero}, {Kepler}, \&
  {Rohrmann}}]{2009ApJ...704.1605A}
{Althaus}, L.~G., {Panei}, J.~A., {Miller Bertolami}, M.~M., {et~al.} 2009,
  \apj, 704, 1605

\bibitem[{{Anderson}(1989)}]{1989ApJ...339..558A}
{Anderson}, L.~S. 1989, \apj, 339, 558

\bibitem[{{Asplund} {et~al.}(2009){Asplund}, {Grevesse}, {Sauval}, \&
  {Scott}}]{2009ARA&A..47..481A}
{Asplund}, M., {Grevesse}, N., {Sauval}, A.~J., \& {Scott}, P. 2009, \araa, 47,
  481

\bibitem[{{Barstow} {et~al.}(2014){Barstow}, {Barstow}, {Casewell}, {Holberg},
  \& {Hubeny}}]{2014MNRAS.440.1607B}
{Barstow}, M.~A., {Barstow}, J.~K., {Casewell}, S.~L., {Holberg}, J.~B., \&
  {Hubeny}, I. 2014, \mnras, 440, 1607

\bibitem[{{Bashkin} \& {Stoner}(1975)}]{1975aelg.book.....B}
{Bashkin}, S. \& {Stoner}, J.~O. 1975, {Atomic energy levels and Grotrian
  Diagrams - Vol.1: Hydrogen I - Phosphorus XV; Vol.2: Sulfur I - Titanium
  XXII}

\bibitem[{{Cartledge} {et~al.}(2005){Cartledge}, {Clayton}, {Gordon},
  {Rachford}, {Draine}, {Martin}, {Mathis}, {Misselt}, {Sofia}, {Whittet}, \&
  {Wolff}}]{2005ApJ...630..355C}
{Cartledge}, S.~I.~B., {Clayton}, G.~C., {Gordon}, K.~D., {et~al.} 2005, \apj,
  630, 355

\bibitem[{{Dere} {et~al.}(1997){Dere}, {Landi}, {Mason}, {Monsignori Fossi}, \&
  {Young}}]{1997A&AS..125..149D}
{Dere}, K.~P., {Landi}, E., {Mason}, H.~E., {Monsignori Fossi}, B.~C., \&
  {Young}, P.~R. 1997, \aaps, 125, 149

\bibitem[{{Drake} {et~al.}(2014){Drake}, {Graham}, {Djorgovski}, {Catelan},
  {Mahabal}, {Torrealba}, {Garc{\'{\i}}a-{\'A}lvarez}, {Donalek}, {Prieto},
  {Williams}, {Larson}, {Christen sen}, {Belokurov}, {Koposov}, {Beshore},
  {Boattini}, {Gibbs}, {Hill}, {Kowalski}, {Johnson}, \&
  {Shelly}}]{2014ApJS..213....9D}
{Drake}, A.~J., {Graham}, M.~J., {Djorgovski}, S.~G., {et~al.} 2014, \apjs,
  213, 9

\bibitem[{{Dreizler} {et~al.}(1995){Dreizler}, {Heber}, {Napiwotzki}, \&
  {Hagen}}]{1995A&A...303L..53D}
{Dreizler}, S., {Heber}, U., {Napiwotzki}, R., \& {Hagen}, H.~J. 1995, \aap,
  303, L53

\bibitem[{{Gianninas} {et~al.}(2010){Gianninas}, {Bergeron}, {Dupuis}, \&
  {Ruiz}}]{2010ApJ...720..581G}
{Gianninas}, A., {Bergeron}, P., {Dupuis}, J., \& {Ruiz}, M.~T. 2010, \apj,
  720, 581

\bibitem[{{Good} {et~al.}(2005){Good}, {Barstow}, {Burleigh}, {Dobbie},
  {Holberg}, \& {Hubeny}}]{2005MNRAS.363..183G}
{Good}, S.~A., {Barstow}, M.~A., {Burleigh}, M.~R., {et~al.} 2005, \mnras, 363,
  183

\bibitem[{{Groote} \& {Hunger}(1976)}]{1976A&A....52..303G}
{Groote}, D. \& {Hunger}, K. 1976, \aap, 52, 303

\bibitem[{{Groote} \& {Hunger}(1977)}]{1977A&A....56..129G}
{Groote}, D. \& {Hunger}, K. 1977, \aap, 56, 129

\bibitem[{{Groote} \& {Hunger}(1997)}]{1997A&A...319..250G}
{Groote}, D. \& {Hunger}, K. 1997, \aap, 319, 250

\bibitem[{{Hamann} {et~al.}(1982){Hamann}, {Sch\"onberner}, \&
  {Heber}}]{1982A&A...116..273H}
{Hamann}, W.-R., {Sch\"onberner}, D., \& {Heber}, U. 1982, \aap, 116, 273

\bibitem[{{Havnes} \& {Goertz}(1984)}]{1984A&A...138..421H}
{Havnes}, O. \& {Goertz}, C.~K. 1984, \aap, 138, 421

\bibitem[{{H{\'e}brard} {et~al.}(2002){H{\'e}brard}, {Friedman}, {Kruk},
  {Lehner}, {Lemoine}, {Linsky}, {Moos}, {Oliveira}, {Sembach}, {Sonneborn},
  {Vidal-Madjar}, \& {Wood}}]{2002P&SS...50.1169H}
{H{\'e}brard}, G., {Friedman}, S.~D., {Kruk}, J.~W., {et~al.} 2002, \planss,
  50, 1169

\bibitem[{{H{\'e}brard} \& {Moos}(2003)}]{2003ApJ...599..297H}
{H{\'e}brard}, G. \& {Moos}, H.~W. 2003, \apj, 599, 297

\bibitem[{{Hesser} {et~al.}(1976){Hesser}, {Walborn}, \&
  {Ugarte}}]{1976Natur.262..116H}
{Hesser}, J.~E., {Walborn}, N.~R., \& {Ugarte}, P.~P. 1976, \nat, 262, 116

\bibitem[{{Hoyer} {et~al.}(2017){Hoyer}, {Rauch}, {Werner}, {Kruk}, \&
  {Quinet}}]{2017A&A...598A.135H}
{Hoyer}, D., {Rauch}, T., {Werner}, K., {Kruk}, J.~W., \& {Quinet}, P. 2017,
  \aap, 598, A135

\bibitem[{{H{\"u}gelmeyer} {et~al.}(2006){H{\"u}gelmeyer}, {Dreizler},
  {Homeier}, {Krzesi{\'n}ski}, {Werner}, {Nitta}, \&
  {Kleinman}}]{2006A&A...454..617H}
{H{\"u}gelmeyer}, S.~D., {Dreizler}, S., {Homeier}, D., {et~al.} 2006, \aap,
  454, 617

\bibitem[{{Jeffery} \& {Hamann}(2010)}]{2010MNRAS.404.1698J}
{Jeffery}, C.~S. \& {Hamann}, W.-R. 2010, \mnras, 404, 1698

\bibitem[{{Krti{\v c}ka} \& {Kub{\'a}t}(2001)}]{2001A&A...377..175K}
{Krti{\v c}ka}, J. \& {Kub{\'a}t}, J. 2001, \aap, 377, 175

\bibitem[{{Krti{\v c}ka} \& {Kub{\'a}t}(2005)}]{2005ASPC..334..337K}
{Krti{\v c}ka}, J. \& {Kub{\'a}t}, J. 2005, in Astronomical Society of the
  Pacific Conference Series, Vol. 334, 14th European Workshop on White Dwarfs,
  ed. D.~{Koester} \& S.~{Moehler}, 337

\bibitem[{{Krti{\v c}ka} {et~al.}(2016){Krti{\v c}ka}, {Kub{\'a}t}, \& {Krti{\v
  c}kov{\'a}}}]{2016A&A...593A.101K}
{Krti{\v c}ka}, J., {Kub{\'a}t}, J., \& {Krti{\v c}kov{\'a}}, I. 2016, \aap,
  593, A101

\bibitem[{{Krti{\v c}ka} {et~al.}(2003){Krti{\v c}ka}, {Owocki}, {Kub{\'a}t},
  {Galloway}, \& {Brown}}]{2003A&A...402..713K}
{Krti{\v c}ka}, J., {Owocki}, S.~P., {Kub{\'a}t}, J., {Galloway}, R.~K., \&
  {Brown}, J.~C. 2003, \aap, 402, 713

\bibitem[{{Krti{\v c}ka} {et~al.}(2010){Krti{\v c}ka}, {Votruba}, \&
  {Kub{\'a}t}}]{2010A&A...516A.100K}
{Krti{\v c}ka}, J., {Votruba}, V., \& {Kub{\'a}t}, J. 2010, \aap, 516, A100

\bibitem[{{Kurucz}(1991)}]{kurucz1991}
{Kurucz}, R.~L. 1991, in NATO ASIC Proc. 341: Stellar Atmospheres - Beyond
  Classical Models, ed. L.~{Crivellari}, I.~{Hubeny}, \& D.~G. {Hummer}, 441

\bibitem[{{Kurucz}(2009)}]{kurucz2009}
{Kurucz}, R.~L. 2009, in American Institute of Physics Conference Series, Vol.
  1171, American Institute of Physics Conference Series, ed. I.~{Hubeny}, J.~M.
  {Stone}, K.~{MacGregor}, \& K.~{Werner}, 43

\bibitem[{{Kurucz}(2011)}]{kurucz2011}
{Kurucz}, R.~L. 2011, Canadian Journal of Physics, 89, 417

\bibitem[{{Landi} {et~al.}(2013){Landi}, {Young}, {Dere}, {Del Zanna}, \&
  {Mason}}]{2013ApJ...763...86L}
{Landi}, E., {Young}, P.~R., {Dere}, K.~P., {Del Zanna}, G., \& {Mason}, H.~E.
  2013, \apj, 763, 86

\bibitem[{{Lemoine} {et~al.}(2002){Lemoine}, {Vidal-Madjar}, {H{\'e}brard},
  {D{\'e}sert}, {Ferlet}, {Lecavelier des {\'E}tangs}, {Howk}, {Andr{\'e}},
  {Blair}, {Friedman}, {Kruk}, {Lacour}, {Moos}, {Sembach}, {Chayer},
  {Jenkins}, {Koester}, {Linsky}, {Wood}, {Oegerle}, {Sonneborn}, \&
  {York}}]{2002ApJS..140...67L}
{Lemoine}, M., {Vidal-Madjar}, A., {H{\'e}brard}, G., {et~al.} 2002, \apjs,
  140, 67

\bibitem[{{Napiwotzki}(1997)}]{1997fbs..conf..207N}
{Napiwotzki}, R. 1997, in The Third Conference on Faint Blue Stars, ed.
  A.~G.~D. {Philip}, J.~{Liebert}, R.~{Saffer}, \& D.~S. {Hayes}, 207

\bibitem[{{Napiwotzki} {et~al.}(2003){Napiwotzki}, {Christlieb}, {Drechsel},
  {Hagen}, {Heber}, {Homeier}, {Karl}, {Koester}, {Leibundgut}, {Marsh},
  {Moehler}, {Nelemans}, {Pauli}, {Reimers}, {Renzini}, \&
  {Yungelson}}]{2003Msngr.112...25N}
{Napiwotzki}, R., {Christlieb}, N., {Drechsel}, H., {et~al.} 2003, The
  Messenger, 112, 25

\bibitem[{{Napiwotzki} \& {Rauch}(1994)}]{1994A&A...285..603N}
{Napiwotzki}, R. \& {Rauch}, T. 1994, \aap, 285, 603

\bibitem[{{Oksala} {et~al.}(2015){Oksala}, {Kochukhov}, {Krti{\v c}ka},
  {Townsend}, {Wade}, {Prv{\'a}k}, {Mikul{\'a}{\v s}ek}, {Silvester}, \&
  {Owocki}}]{2015MNRAS.451.2015O}
{Oksala}, M.~E., {Kochukhov}, O., {Krti{\v c}ka}, J., {et~al.} 2015, \mnras,
  451, 2015

\bibitem[{{Pedersen} \& {Thomsen}(1977)}]{1977A&AS...30...11P}
{Pedersen}, H. \& {Thomsen}, B. 1977, \aaps, 30, 11

\bibitem[{{Rauch} \& {Deetjen}(2003)}]{2003ASPC..288..103R}
{Rauch}, T. \& {Deetjen}, J.~L. 2003, in Astronomical Society of the Pacific
  Conference Series, Vol. 288, Stellar Atmosphere Modeling, ed. I.~{Hubeny},
  D.~{Mihalas}, \& K.~{Werner}, 103

\bibitem[{{Rauch} {et~al.}(2017){Rauch}, {Gamrath}, {Quinet}, {Hoyer},
  {Werner}, {Quinet}, {Kruk}, \& {Demleitner}}]{2017rauch}
{Rauch}, T., {Gamrath}, S., {Quinet}, P.~{L\"obling}, L., {et~al.} 2017, \aap,
  599, A142

\bibitem[{{Reindl} {et~al.}(2014{\natexlab{a}}){Reindl}, {Rauch}, {Werner},
  {Kepler}, {G{\"a}nsicke}, \& {Gentile Fusillo}}]{2014A&A...572A.117R}
{Reindl}, N., {Rauch}, T., {Werner}, K., {et~al.} 2014{\natexlab{a}}, \aap,
  572, A117

\bibitem[{{Reindl} {et~al.}(2014{\natexlab{b}}){Reindl}, {Rauch}, {Werner},
  {Kruk}, \& {Todt}}]{2014A&A...566A.116R}
{Reindl}, N., {Rauch}, T., {Werner}, K., {Kruk}, J.~W., \& {Todt}, H.
  2014{\natexlab{b}}, \aap, 566, A116

\bibitem[{{Seaton} {et~al.}(1994){Seaton}, {Yan}, {Mihalas}, \&
  {Pradhan}}]{1994MNRAS.266..805S}
{Seaton}, M.~J., {Yan}, Y., {Mihalas}, D., \& {Pradhan}, A.~K. 1994, \mnras,
  266, 805

\bibitem[{{Smith} \& {Bohlender}(2007)}]{2007A&A...475.1027S}
{Smith}, M.~A. \& {Bohlender}, D.~A. 2007, \aap, 475, 1027

\bibitem[{{Sofia} {et~al.}(2005){Sofia}, {Wolff}, {Rachford}, {Gordon},
  {Clayton}, {Cartledge}, {Martin}, {Draine}, {Mathis}, {Snow}, \&
  {Whittet}}]{2005ApJ...625..167S}
{Sofia}, U.~J., {Wolff}, M.~J., {Rachford}, B., {et~al.} 2005, \apj, 625, 167

\bibitem[{{Springmann} \& {Pauldrach}(1992)}]{1992A&A...262..515S}
{Springmann}, U.~W.~E. \& {Pauldrach}, A.~W.~A. 1992, \aap, 262, 515

\bibitem[{{Unglaub}(2006)}]{2006BaltA..15..147U}
{Unglaub}, K. 2006, Baltic Astronomy, 15, 147

\bibitem[{{Unglaub}(2008)}]{2008A&A...486..923U}
{Unglaub}, K. 2008, \aap, 486, 923

\bibitem[{{Unglaub} \& {Bues}(2000)}]{2000A&A...359.1042U}
{Unglaub}, K. \& {Bues}, I. 2000, \aap, 359, 1042

\bibitem[{{Voigt}(2012)}]{voigt...abriss}
{Voigt}, H.-H. 2012, {Abriss der Astronomie. 6. Auflage, Wiley-VCH}

\bibitem[{{Werner}(1996)}]{1996ApJ...457L..39W}
{Werner}, K. 1996, \apjl, 457, L39

\bibitem[{{Werner} {et~al.}(2003{\natexlab{a}}){Werner}, {Deetjen}, {Dreizler},
  {Nagel}, {Rauch}, \& {Schuh}}]{2003ASPC..288...31W}
{Werner}, K., {Deetjen}, J.~L., {Dreizler}, S., {et~al.} 2003{\natexlab{a}}, in
  Astronomical Society of the Pacific Conference Series, Vol. 288, Stellar
  Atmosphere Modeling, ed. I.~{Hubeny}, D.~{Mihalas}, \& K.~{Werner}, 31

\bibitem[{{Werner} \& {Dreizler}(1999)}]{1999JCoAM.109...65W}
{Werner}, K. \& {Dreizler}, S. 1999, Journal of Computational and Applied
  Mathematics, 109, 65

\bibitem[{{Werner} {et~al.}(1997{\natexlab{a}}){Werner}, {Dreizler}, {Heber},
  \& {Rauch}}]{1997ASSL..214..207W}
{Werner}, K., {Dreizler}, S., {Heber}, U., \& {Rauch}, T. 1997{\natexlab{a}},
  in Astrophysics and Space Science Library, Vol. 214, White dwarfs, ed.
  J.~{Isern}, M.~{Hernanz}, \& E.~{Garcia-Berro}, 207

\bibitem[{{Werner} {et~al.}(1995){Werner}, {Dreizler}, {Heber}, {Rauch},
  {Wisotzki}, \& {Hagen}}]{1995A&A...293L..75W}
{Werner}, K., {Dreizler}, S., {Heber}, U., {et~al.} 1995, \aap, 293, L75

\bibitem[{{Werner} {et~al.}(2003{\natexlab{b}}){Werner}, {Dreizler}, {Kruk}, \&
  {Sitko}}]{2003whdw.conf..171W}
{Werner}, K., {Dreizler}, S., {Kruk}, J.~W., \& {Sitko}, M.~L.
  2003{\natexlab{b}}, in NATO ASIB Proc. 105: White Dwarfs, ed. D.~{de
  Martino}, R.~{Silvotti}, J.-E. {Solheim}, \& R.~{Kalytis}, Vol. 105, 171

\bibitem[{{Werner} {et~al.}(2012){Werner}, {Dreizler}, \& {Rauch}}]{tmap2012}
{Werner}, K., {Dreizler}, S., \& {Rauch}, T. 2012, {TMAP: T{\"u}bingen NLTE
  Model-Atmosphere Package}, Astrophysics Source Code Library

\bibitem[{{Werner} {et~al.}(1999){Werner}, {Dreizler}, {Rauch}, {Barnstedt},
  {G{\"o}lz}, {Gringel}, {Kappelmann}, {Kr{\"a}mer}, {Widmann}, {Koesterke},
  {Haas}, {Heber}, {Appenzeller}, \& {Grewing}}]{1999ASPC..169..511W}
{Werner}, K., {Dreizler}, S., {Rauch}, T., {et~al.} 1999, in Astronomical
  Society of the Pacific Conference Series, Vol. 169, 11th European Workshop on
  White Dwarfs, ed. S.-E. {Solheim} \& E.~G. {Meistas}, 511

\bibitem[{{Werner} {et~al.}(1997{\natexlab{b}}){Werner}, {Dreizler}, {Rauch},
  \& {Heber}}]{1997fbs..conf..227W}
{Werner}, K., {Dreizler}, S., {Rauch}, T., \& {Heber}, U. 1997{\natexlab{b}},
  in The Third Conference on Faint Blue Stars, ed. A.~G.~D. {Philip},
  J.~{Liebert}, R.~{Saffer}, \& D.~S. {Hayes}, 227

\bibitem[{{Werner} {et~al.}(2014){Werner}, {Rauch}, \&
  {Kepler}}]{2014A&A...564A..53W}
{Werner}, K., {Rauch}, T., \& {Kepler}, S.~O. 2014, \aap, 564, A53

\bibitem[{{Werner} {et~al.}(2015){Werner}, {Rauch}, \&
  {Kruk}}]{2015A&A...582A..94W}
{Werner}, K., {Rauch}, T., \& {Kruk}, J.~W. 2015, \aap, 582, A94

\bibitem[{{Werner} {et~al.}(2017){Werner}, {Rauch}, \&
  {Kruk}}]{2017A&A...601A...8W}
{Werner}, K., {Rauch}, T., \& {Kruk}, J.~W. 2017, \aap, 601, A8

\bibitem[{{Ziegler} {et~al.}(2012){Ziegler}, {Rauch}, {Werner}, {K{\"o}ppen},
  \& {Kruk}}]{2012A&A...548A.109Z}
{Ziegler}, M., {Rauch}, T., {Werner}, K., {K{\"o}ppen}, J., \& {Kruk}, J.~W.
  2012, \aap, 548, A109

\end{thebibliography}

\end{document}